\documentstyle[12pt,preprint]{aastex}  

\begin{document}

                           % The preamble begins here.
\title{2007 TY430: A Cold Classical Kuiper Belt Type Binary in the Plutino Population}
\author{Scott S. Sheppard\altaffilmark{1}, Darin Ragozzine\altaffilmark{2} and Chadwick Trujillo\altaffilmark{3}}

\altaffiltext{1}{Department of Terrestrial Magnetism, Carnegie Institution of Washington, 5241 Broad Branch Rd. NW, Washington, DC 20015, USA, sheppard@dtm.ciw.edu}
\altaffiltext{2}{Harvard-Smithsonian Astrophysical Observatory, Cambridge, MA 02138, USA}
\altaffiltext{3}{Gemini Observatory, 670 North A'ohoku Place, Hilo, HI 96720, USA}

\begin{abstract}  % Produces abstract

Kuiper Belt object 2007 TY430 is the first wide, equal-sized, binary
known in the 3:2 mean motion resonance with Neptune. The two
components have a maximum separation of about 1 arcsecond and are on
average less than 0.1 magnitudes different in apparent magnitude with
identical ultra-red colors ($g-i=1.49\pm0.01$ mags).  Using nearly
monthly observations of 2007 TY430 from 2007-2011, the orbit of the
mutual components was found to have a period of $961.2\pm4.6$ days
with a semi-major axis of $21000\pm160$ km and eccentricity of
$0.1529\pm0.0028$.  The inclination with respect to the ecliptic is
$15.68\pm0.22$ degrees and extensive observations have allowed the
mirror orbit to be eliminated as a possibility.  The total mass for
the binary system was found to be $7.90\pm0.21 \times10^{17}$ kg.
Equal-sized, wide binaries and ultra-red colors are common in the low
inclination ``cold'' classical part of the Kuiper Belt and likely
formed through some sort of three body interactions within a much
denser Kuiper Belt.  To date 2007 TY430 is the only ultra-red,
equal-sized binary known outside of the classical Kuiper belt
population.  Numerical simulations suggest 2007 TY430 is moderately
unstable in the outer part of the 3:2 resonance and thus 2007 TY430 is
likely an escaped ``cold'' classical object that later got trapped in
the 3:2 resonance.  Similar to the known equal-sized, wide binaries in
the cold classical population, the binary 2007 TY430 requires a high
albedo and very low density structure to obtain the total mass found
for the pair.  For a realistic minimum density of 0.5 g/cm$^{3}$ the
albedo of 2007 TY430 would be greater than 0.17.  For reasonable
densities, the radii of either component should be less than 60 km,
and thus the relatively low eccentricity of the binary is interesting
since no tides should be operating on the bodies at their large
distances from each other.  The low prograde inclination of the binary
also makes it unlikely the Kozai mechanism could have altered the
orbit, making the 2007 TY430 binary orbit likely one of the few
relatively unaltered primordial binary orbits known.  Under some
binary formation models, the low inclination prograde orbit of the
2007 TY430 binary indicates formation within a relatively high
velocity regime in the Kuiper Belt.

\end{abstract}


\section{Introduction}

Small solar system bodies, such as asteroids and Kuiper Belt Objects
(KBOs), may be leftover remnants of the planetesimals that went into
the formation of the terrestrial and giant planets.  These small
bodies' orbital and physical properties give insight into the Solar
System's origins and evolution.  KBO binaries are particularly
informative about the conditions at the time when they formed, which
must have been at an earlier epoch of Solar System history.

\begin{figure}
\epsscale{0.4}
\centerline{\includegraphics[angle=90,totalheight=0.6\textheight]{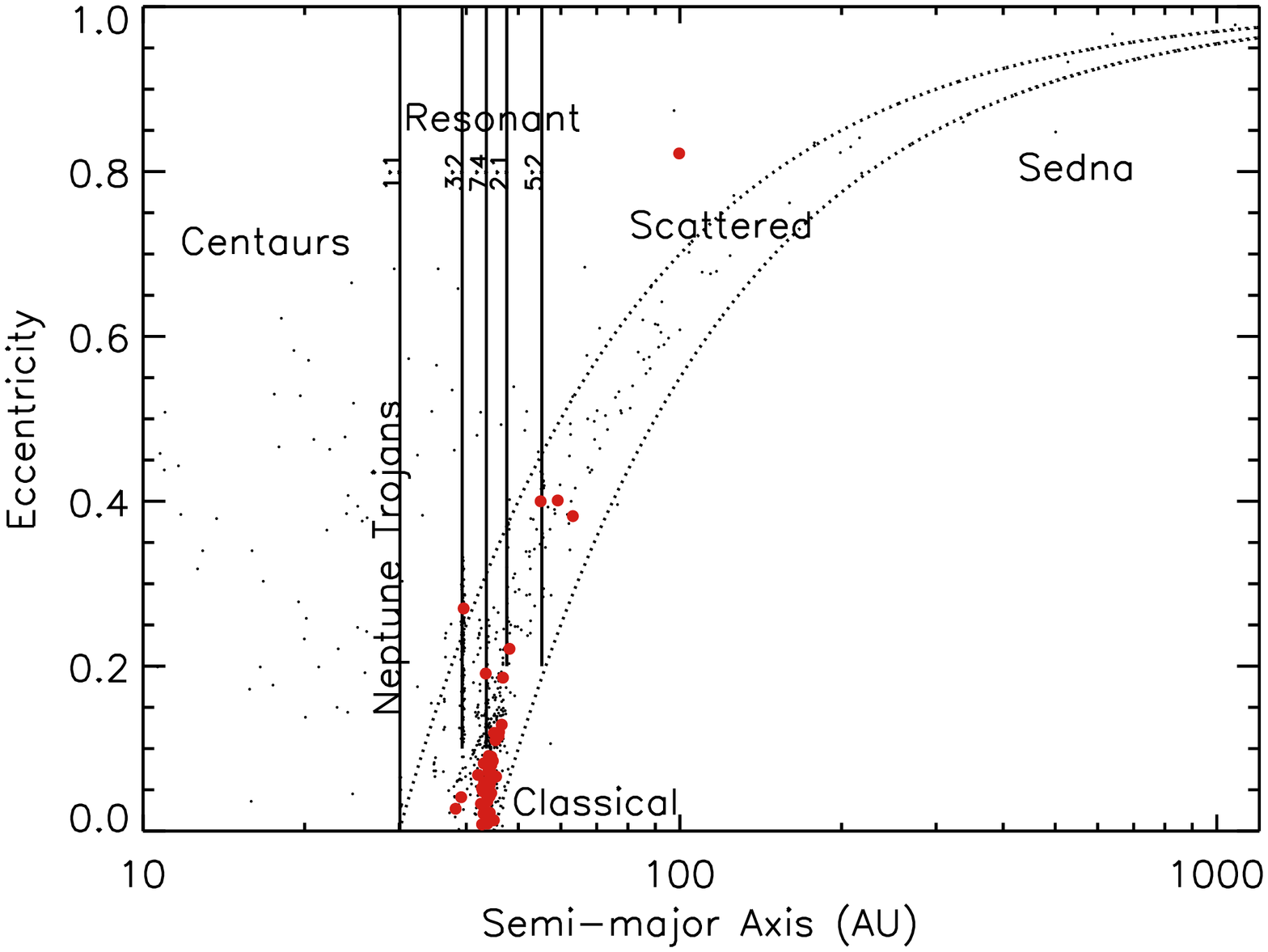}}
\caption{The semi-major axis versus eccentricity for all
  multi-opposition observed TNOs.  Objects considered equal-sized
  binaries ($\Delta$mag$<1$ mag) are shown with big filled circles.
  This figure shows several distinct dynamical KBO
  populations. Vertical dashed lines show the main resonances with
  Neptune as well as the Neptune Trojans in the $1:1$ resonance.
  Scattered disk objects have perihelia $30 \lesssim q \lesssim 45$
  AU as shown between the dashed lines.  Classical objects are in the
  lower center portion of the figure and include the the Main Kuiper
  Belt (MKB) with its high and low inclination populations.  An edge
  near 50 AU can clearly be seen for low eccentricity objects.
  Centaurs are on unstable orbits between the giant planets.}
\label{fig:kboeabinary} 
\end{figure}

\begin{figure}
\epsscale{0.4}
\centerline{\includegraphics[angle=90,totalheight=0.6\textheight]{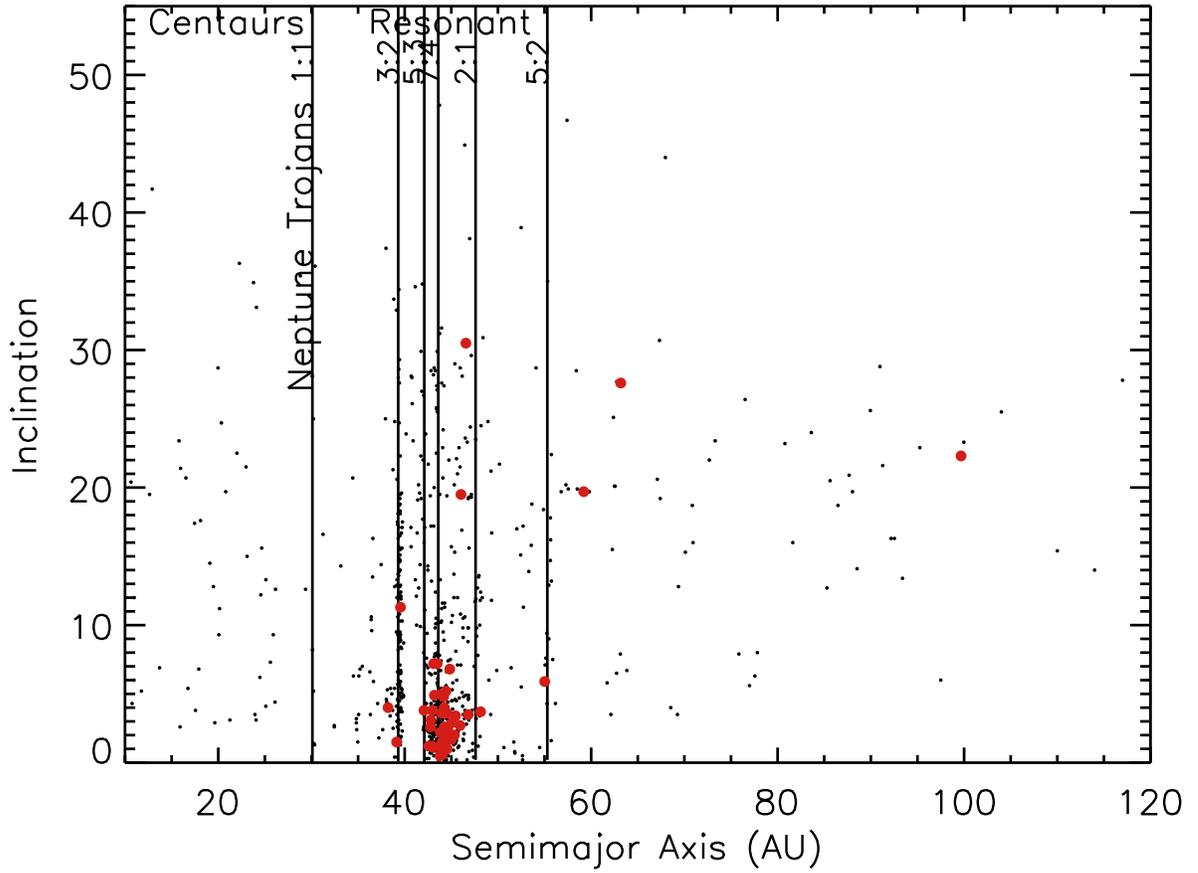}}
\caption{The semi-major axis versus inclination for all
  multi-opposition observed TNOs.  Objects considered equal-sized
  binaries ($\Delta$mag$<1$ mag) are shown with big filled circles.}
\label{fig:kboiabinary} 
\end{figure}

The Kuiper Belt is dynamically structured with three main dynamical
classes (Figures~\ref{fig:kboeabinary}
to~\ref{fig:kboiabinaryblowup}).  1) Classical KBOs have semi-major
axes between about 40 and 50 AU with moderate eccentricities and
inclinations.  These objects may be regarded as the population
originally predicted for the Kuiper Belt (Fernandez and Ip 1981), but
they have higher eccentricities and inclinations than expected (Jewitt
et al. 1998).  The dynamics of the classical KBOs have shown that the
outer Solar System has been highly modified through the evolution of
the planets (Hahn and Malhotra 2005).  There appears to be two subsets
of classical KBOs.  The low inclination ``cold'' classical population
has generally smaller and redder objects compared to the high
inclination ``hot'' classical population (Tegler and Romanishin 2000;
Brown 2001; Levison and Stern 2001; Trujillo and Brown 2002; Peixinho
et al. 2008).  These two classical populations likely represent
different formation regions from the outer solar system.  Through
migration and scattering of the planets very early in the Solar System
the two classical populations came to reside where we see them now
(Gomes 2003; Levison and Morbidelli 2003; Batygin et al. 2011). 2)
Scattered disk objects have large eccentricities with perihelia near
the orbit of Neptune ($q \sim 30-45$ AU).  The scattered disk objects
are likely to have been moved to their current orbits through
interactions with Neptune (Gomes et al. 2008).

%Dawson and Murray-Clay 2011.

3) Resonant KBOs are in mean motion resonances with Neptune and
generally have higher eccentricities and inclinations than classical
KBOs.  Most of the resonance objects were likely captured into their
resonances from the outward migration and circularization of Neptune's
orbit (Malhotra 1995; Levison et al. 2008).  Where the resonance
objects came from and how and when this capture occurred is still
highly debated.  The 3:2 resonance, called Plutinos since Pluto is in
this resonance, appears to be the most populated resonance (Jewitt et
al. 1998; Chiang and Jordan 2002; Sheppard et al. 2011).  The other
resonances with a significant number of known objects are the outer
mean motion resonances 5:3, 7:4, 2:1 and 5:2 (Elliot et al. 2005; Hahn
and Malhotra 2005; Gladman et al. 2008).

Binary small bodies are a particularly good way to obtain clues about
the Solar System's past.  Remarkably, the small bodies in the various
Solar System reservoirs appear to have significantly different binary
characteristics (Richardson and Walsh 2006; Noll et al. 2008a).  In
the main asteroid belt, most secondaries are close to and
significantly smaller than the primaries (Merline et al. 2002; Margot
2002; Marchis et al. 2006b).  This is indicative of formation through
direct collisions.  The few binaries known in the Jupiter Trojan
population hint at these objects being of very low density, implying
they may be similar to comets and Trans-Neptunian Objects (TNOs)
(Marchis et al. 2006a).  Starting with the discovery of the first KBO
binary other than Pluto, it has become apparent that most of the low
inclination classical Kuiper Belt binaries have large separations and
similar sized components (Veillet et al. 2002; Noll et al. 2008a).
Because of the large angular momentum in these equal-sized, wide
binary systems, they cannot be formed by direct collisions.  Direct
formation by gravitational collapse, the dynamical friction of a sea
of KBOs, three body interactions, some other collisionless
interactions, or a combination of these are the best viable scenarios
for binary formation in the classical Kuiper Belt (Nesvorn{\'y} et
al. 2010; Goldreich et al. 2002; Weidenschilling 2002; Funato et
al. 2004; Astakhov et al. 2005; Lee 2007; Schlichting and Sari 2008a;
Gamboa Suarez et al. 2010).  The cold classical disk also has a high
fraction of binaries, irrespective of separation and difference in
magnitude of the individual components, compared to the hot classical,
resonant and scattered populations, about 30 percent versus 5 percent
binary fraction, respectively (Stephens and Noll 2006; Noll et
al. 2008b).

Unlike the low inclination classical KBOs, many of the other known KBO
binaries do not have such equal-sized and wide components.  The
different types of binary formation mechanisms proposed will each tend
to create a particular subset of satellite semi-major axes,
eccentricities, inclinations and secondary diameters (Kern and Elliot
2006a; Noll et al. 2008a).  Thus, understanding the orbital elements
and physical characteristics of the components of a binary is
essential to constraining how the binary may have formed.
Understanding the formation mechanism of binaries gives insight into
the original solar nebula and the collisional environment in the
distant past.  The two distinct binary populations seen in the cold
and hot dynamical classes are strong evidence of a different
history. For example, Parker and Kavelaars (2010) use the fragile
nature of the wide binaries in the cold classical region as evidence
to show that cold classical KBOs were not likely to have been
implanted dynamically, as suggested by Levison et al. (2008). Binaries
also likely played an important role in planet formation by acting as
a heat source in the planetesimal disk, tranforming gravitational
potential energy into kinetic energy of planetesimals in the disk
(Perets 2011).

2007 TY430 is a unique binary KBO because it is the first known
equal-sized, wide binary in the 3:2 mean motion resonance with
Neptune, which is the innermost well populated Neptune resonance.  The
other known 3:2 resonance binary objects (including Pluto, Orcus,
(47171) 1999 TC36 and (208996) 2003 AZ84) have significantly smaller
and relatively closer secondaries (magnitude differences greater than
2 mags) that are more indicative of formation through direct
collisions (Stern 2002; Brown et al. 2006; Canup 2011).  Because 2007
TY430 indicates a different binary formation mechanism operating
within the 3:2 resonance population, 2007 TY430 was extensively
observed over the past several years in order to determine the orbital
and physical properties of its two components.

\section{Observations}

KBO 2007 TY430 was discovered with the 8.2 meter Subaru telescope
(Sheppard et al. 2008) during a survey for faint Neptune Trojans
(Sheppard and Trujillo 2010a).  The Suprime-Cam wide-field imager was
used, which has ten $2048\times4096$ pixel CCDs with a pixel scale of
$0.20$ arcseconds per pixel (Miyazaki et al. 2002).  At discovery it
was apparent that 2007 TY430 was elongated compared to the background
stars.  Upon closer examination it was found that 2007 TY430 was a
binary with the two components having a separation of about 0.67
arcseconds and within about 0.1 magnitudes of each other (Sheppard and
Trujillo 2008).  2007 TY430 was at a heliocentric distance of about
29.3 AU when discovered.  Table 1 shows the heliocentric orbit
determined for 2007 TY430.

Director's Discretionary Time (DDT) was immediately requested with the
8.1 meter Gemini telescope in order to monitor the binary to obtain
the orbit of the mutual components (Figure~\ref{fig:image2007TY430}).
GMOS on Gemini South was used for the 2007 to early 2008 DDT
observations.  GMOS queue time on Gemini North was used for the later
2008 through 2011 observations.  Each GMOS has three $2048\times 4608$
pixel CCDs with 0.0730 arcseconds per pixel for GMOS South and 0.0728
arcseconds per pixel for GMOS North (Hook et al. 2004).  The
observations were only executed in excellent seeing ($<0.6$ arcseconds
FWHM).  Mainly Sloan i-band images were obtained because of the
superior seeing at longer wavelengths from the ground, but some Sloan
g and r-band images were obtained when the two components of the
binary were well separated to measure their individual colors.

Table 2 shows the astrometry and photometry between the two components
of the binary for all observations obtained for 2007 TY430.  In
addition to the Subaru and Gemini data, 2007 TY430 was observed with
the Wide Field Planetary Camera 2 on the Hubble Space Telescope in
September 2008 and is included in Table 2 (Noll et al. 2009).  The
maximum observed separation of the two components was about one
arcsecond.  The maximum difference in magnitude between the components
was about 0.3 mags.  Many times the two components had nearly equal
magnitudes with an average difference of about 0.1 mags.  A few times,
the two components switched as to which one was brighter (Table 2).
The relatively fast change in brightness differences between the two
components compared to the long orbital period of the binary, and the
differences in brightness when observed near the same spot in the
orbit for two different epochs, show that the individual components
are not in synchronous rotation.

\section{Analysis}

\subsection{Heliocentric Orbit of 2007 TY430 and the 3:2 Resonance}

Table 1 shows the heliocentric orbital elements for 2007 TY430 from
the minor planet center.  The semi-major axis and eccentricity are
easily recognized as a 3:2 resonance object or Plutino.
Figure~\ref{fig:kboeabinary} shows 2007 TY430 is near the outer edge
of the 3:2 resonance.  In order to determine the stability of 2007
TY430's orbit, 10 clones were made using the orbital elements and full
covariance matrix of the orbital elements from the AstDys website
(http://hamilton.dm.unipi.it/astdys/) including observations spanning
a more than 3 year arc.  Using the Swift integrator program (Levison
and Duncan 2000) with the four giant planets and relatively short time
steps of about 40 days, the clones were numerically integrated
backwards in time for the age of the Solar System.  All ten clones are
stable in the 3:2 resonance for the first 1 Myrs.  Eight of the ten
clones remained in the 3:2 resonance for at least 10 Myrs, which based
on the classification method of Gladman et al. (2008), suggests that
it is acceptable to call 2007 TY430 a Plutino, but likely an unstable
one.  Four of the ten remain in the 3:2 resonance for the age of the
Solar System with some interactions with the Kozai resonance.

The loss of some of the 2007 TY430 clones is not surprising since
objects in the 3:2 resonance can be unstable over the age of the solar
system (de Elia et al. 2008; Almeida et al. 2009; di Sisto et
al. 2010).  2007 TY430 has a large libration amplitude of $\sim$120
degrees (see also orbit calculations from Marc Buie's website at
www.boulder.swri.edu/buie/kbo/astrom that has up to date information
first published in Elliot et al. 2005).  For the eccentricity and
inclination of 2007 TY430, this libration amplitude puts it in the
unstable part of the Plutino resonance (Tiscareno and Malhotra 2009),
so it is not suprising that the orbital integrations showed a lack of
long-term stability.

Since the backward numerical integrations did not show 2007 TY430 to
be a strongly stable object, its past history cannot be reliably
constructed. There are two major possibilities for the origin of 2007
TY430. The first is that 2007 TY430 was swept up into the 3:2
resonance, which is the likely case for most of the Plutinos (Hahn and
Malhotra 2005).  These Plutinos likely formed closer to the Sun than
their current locations.  Based on the Brouwer's constant as described
in Murray-Clay and Schlichting (2011), the current relatively large
heliocentric orbital eccentricity of 2007 TY430 (Table 1) suggests that
if it was captured into the 3:2 resonance by an outward, smooth, slowly
migrating Neptune, 2007 TY430 would originally have had a semi-major
axis of about 31 AU if initially in a circular orbit.  Over time,
chaotic diffusion could have caused 2007 TY430 to move to higher and
higher libration amplitudes until reaching its current situation of
being near the outer ``edge'' of the 3:2 resonance. The second
possibility is that 2007 TY430 used to be in a different region of the
Kuiper belt (e.g., the low inclination ``cold'' classical region) and
escaped into the scattered disk.  While in the scattered disk, 2007
TY430 could have entered into the outskirts of the 3:2 resonance in
the common phenomenon of resonance sticking (Lykawka and Mukai 2007),
as the 3:2 resonance is known to catch scattered-disk objects (Lykawka
and Mukai 2005). Resonance sticking naturally explains the large
libration amplitude at the present epoch and thus, dynamically, the
second scenario above seems the most likely.  If scenario two is
correct, it is unknown if 2007 TY430 came from an inner extension of
what is now the cold classical belt or from what we see as the
classical belt today before being pushed into the resonance. How long
2007 TY430 has been in the Plutino resonance and how long ago it left
the cold classical region also cannot be determined.

\subsection{Orbit of Mutual Components of 2007 TY430}

Using a two-body version of the code developed in Ragozzine and Brown
(2009), the best-fit orbital properties of 2007 TY430 were found using
thousands of individual Powell and Levenburg-Marquardt minimizations
(Table 3). The combination of these minimizations provides a strong
quadratic locus of points in chi-squared ($\chi^2$) vs. parameter
space, as expected, with clear evidence that the global minimum was
found.  A single Keplerian orbit provides an excellent fit (reduced
chi-squared ($\chi^2_{red}) = 0.74$ with 46 degrees of freedom) and
thus no non-Keplerian parameters were included
(Figure~\ref{fig:orbitfit}).  The residuals to the fit follow a
Gaussian distribution. The data span a large enough range in time and
viewing angle that the mirror degeneracy (which has $\chi^2_{red}
\simeq 2$) is strongly broken and the mirror orbital solution is
clearly rejected.  The season of potential mutual events ended several
years ago.

Following the method of Brown et al. (2005) from Press et al. (1992),
a Monte-Carlo suite of 29 random realizations of the data was created
and the global minimum for each was determined for each data set. The
best-fit for each fit parameter and ancillary value are computed and
then the standard deviation of the results is reported as the error in
Table 3. Investigation of the distributions of the parameters showed
that all the errors were nearly Gaussian. The mean and median of the
parameters retrieved from the Monte-Carlo simulation were not
significantly different from the best-fit value, therefore the
best-fit on the nominal data-set is listed as the central estimate of
each parameter as shown in Table 3.

From the parameters in Table 3, the current obliquity is $25 \pm 0.5$
degrees. With the range of heliocentric inclinations observed from the
stable clones (7-18 degrees), the mutual inclination never reaches
above about 30 degrees, suggesting that the 2007 TY430 binary in its
current heliocentric orbit would not be subject to Kozai oscillations
(Perets and Naoz 2009, Ragozzine 2009).

\subsection{Physical Parameters of 2007 TY430}

The colors of 2007 TY430 are shown in Table 4.  2007 TY430 is one of
the reddest known objects ever observed ($g-i=1.49\pm0.01$ mags) and
is well above the ultra-red color cutoff ($g-i > 1.2$ mags) defined in
Sheppard (2010).  The term ultra-red objects was coined by Jewitt
(2002) and has become associated with the low inclination classical
Kuiper belt (Tegler and Romanishin 2000; Trujillo and Brown 2002;
Peixinho et al. 2008) or Oort cloud objects (Sheppard 2010) because
most objects in these regions appear to have ultra-red surface colors.
The composition of the ultra-red material is unknown, but spectral
synthesis analysis suggests that the ultra-red color may be associated
with organic material (Doressoundiram et al. 2008; Barucci et
al. 2008; Fulchignoni et al. 2008).

Because of the near linearity in the optical colors of most KBOs
(Doressoundiram et al. 2008), the spectral gradient, $S$, of an object
can be found using two unique optical broad band filters.  The
spectral gradient is basically a very low resolution spectrum of an
object and is usually expressed in percent of reddening per 100 nm in
wavelength.  Following Sheppard (2010), the spectral gradient for 2007
TY430 was found to be $S=36\pm2$, putting 2007 TY430 within the low
inclination ``cold'' classical Kuiper Belt color population shown in
Table 5 of Sheppard (2010).

Both components of the 2007 TY430 binary have indistinguishable
ultra-red colors (Table 4).  Like 2007 TY430, the colors of each
component of other TNO binary pairs are similar to each other, though
the colors of each binary system as a whole span a wide range of
colors from near neutral to ultra-red.  This suggests formation of
each TNO binary in a specific location within a homogeneous disk
during formation of the Solar System (Benecchi et al. 2009).  Infrared
observations of 2007 TY430 using the Gemini NIRI instrument with the
narrow band water and methane filters (Trujillo et al. 2011) show 2007
TY430 does not have any significant amounts of water or methane ice on
its surface (Table 4).

Similar to the moderate to high albedos found for low inclination cold
classical objects (Brucker et al. 2009), the equal-sized, wide,
ultra-red binary 2007 TY430 requires a moderate to high albedo and
very low density structure to obtain the total mass found for the pair
(Figure~\ref{fig:densityradius}).  For a likely absolute minimum
density of 0.5 g/cm$^{3}$, the albedo of 2007 TY430 would be greater
than 0.17.  For reasonable densities, the radii of either component
should be less than 60 km, and thus the relatively low eccentricity of
the binary is interesting since no tides should be operating on the
bodies at their large distances from each other.  For instance,
assuming a realistic density of 0.75 g/cm$^{3}$, results in an albedo
of about 0.23 and radii of the individual components of about 50 km.

\subsection{Percentage of Wide Binaries in Kuiper Belt}

Through an ultra-deep survey for Neptune Trojans using the Subaru and
Magellan telescopes (Sheppard and Trujillo 2010a, 2010b), only one
object, 2007 TY430, was found to be a binary. The Neptune Trojan
ultra-deep survey detected about 900 faint KBOs.  Interestingly, the
brightest and likely one of the largest objects detected in the
ultra-deep survey was 2007 TY430 ($m_{R}=20.94\pm0.01$).  The vast
majority of the objects found in the Neptune Trojan survey were very
faint ($m_{R}\sim 24-25$ mag) KBOs with expected radii of only a few
tens of kilometers.  These discoveries were all checked by eye for
binarity at discovery, like 2007 TY430.  Most were not followed up to
obtain orbits, but the survey was conducted within a few degrees of
the ecliptic with most KBOs found to be between 40 and 50 AU, in the
heart of the cold classical belt.  Adding in the hundred or so KBOs
detected in a similar ultra-deep survey for satellites of Saturn,
Uranus and Neptune (Sheppard et al. 2005; 2006), puts the number of
KBOs detected in excellent seeing conditions at about one thousand
with only one binary detection ($0.1^{+0.1}_{-0.07}$ percent).  The
vast majority of the KBOs discovered in the Neptune Trojan survey were
in excellent seeing ranging between 0.45 and 0.8 arcseconds, allowing
easy detection of any found widely separated binaries.  Only the
brightest KBO found showed a binary nature.  Nesvorn{\'y} et
al. (2011) demonstrated that binaries that may have formed through
gravitational collapse with radii of less than or equal to 2007
TY430's are less likely to survive because collisions should be able
to disrupt the pair over time.  This is consistent with the
non-detection of any binaries for the thousand smaller KBOs imaged in
very good seeing conditions, though observational biases must be taken
into account (Naoz et al. 2010).

Previous ground based surveys have estimated about one percent of TNOs
to be widely separated equal-sized binaries.  These surveys
detected significantly brighter and thus likely larger objects.  The
Deep Ecliptic Survey (DES), with an average photometric depth of only
22.5 magnitudes in the R-band (Elliot et al. 2005), found 4 out of 634
objects to be binaries from the ground, of which only about 212 were
found in good seeing with a fine pixel scale using the Magellan
telescope (Osip et al. 2003; Kern and Elliot 2006a).  An unpublished
Keck survey targeted 150 known KBOs and found no obvious binaries from
the ground, but it is unknown the brightness distribution of the
objects observed (Schaller and Brown 2003).  Lin et al. (2010) found
two nearly equal mass binaries in the Canada-France ecliptic plane
survey and from this suggest that over 1.5 percent of low inclination
classical Kuiper Belt objects likely have large separation equal-sized
binaries ($>0.4$ arcsecond separation).  Thus surveys of brighter KBOs
show that about 1 to 2 percent are binaries with separations
detectable from the ground, while this work shows that there is likely
significantly less than this for fainter and smaller objects.

There are observational biases against detecting binaries at smaller
sizes.  For instance, the Hill Spheres (see section 4.1) of the
fainter and thus smaller objects are smaller, making any stable
binaries closer to each other.  This work went over two magnitudes
fainter than the DES, thus it was sensitive to objects that were a
factor of 2.5 smaller, with a factor 2.5 smaller Hill radius.  This
smaller Hill radius would have made the 2007 TY430 binary undetectable
from the ground at discovery, but it would still have been an obvious
binary if imaged when the components were their furthest distance from
each other, which is the configuration in which eccentric binaries
spend most of their time.  From Table 5, four wide binaries would
still have been detected from the ground even if their separations
were halved, as their separations at discovery would have been about
0.6 arcseconds or greater, similar to the separation of 2007 TY430 at
discovery.  Observing many very faint KBOs with HST, in a
serendipitous (such as Fuentes et al. 2010) as well as a direct
targeted search, would help eliminate this selection effect and help
determine if smaller objects really do have fewer satellites than
larger objects.

\section{Discussion}

The ultra-red color and wide equal-sized binary nature of 2007 TY430
makes it very similar to the unique characteristics of the low
inclination classical Kuiper Belt.  This strongly favors the argument
that 2007 TY430 originated as a typical low inclination ``cold''
classical Kuiper Belt object (assuming that the cold classical
population forms in situ).  Weak chaos over long timescales (or
stronger chaos on faster timescales) slowly removed 2007 TY430 from
the cold classical region into a dynamically excited orbit in the
scattered disk (see Volk and Malhotra 2011).  Eventually 2007 TY430
became trapped in the unstable outskirts of the 3:2 resonance where it
is seen today. During this process, the binary remained stable.  This
would suggest that 2007 TY430's formation environment is likely the
cause of its ultra-red color and wide binary nature and not its
current orbital characteristics. This bolsters the argument of
Benecchi et al. (2009) who suggest that similar colors for binary
components indicate common formation conditions, regardless of present
location.

\subsection{Stability of 2007 TY430}

The Hill sphere of an object orbiting the Sun is the region where
satellites are generally stable around the object, given by:

\begin{equation}
r_{H} = a_{p} \left[\frac{(m_{p1}+m_{p2})}{3M_{\odot}}\right]^{1/3}
\label{eq:hill}
\end{equation}

where $a_{p}$ is heliocentric distance of the object in AU,
$M_{\odot}$ is the mass of the Sun and $m_{p1}$ and $m_{p2}$ are the
mass of the first and second components of the binary, respectively.
The Hill radius for 2007 TY430 is about $3.015\times 10^{5}$ km.  

Using the heliocentric semi-major axis of 2007 TY430 from Table 1 and
a total mass of $7.90\pm0.21 \times10^{17}$ kg for the binary, gives
$a_{bin}/r_{H}$ of about 0.07 for 2007 TY430.  While this is still
well within the stable region of orbital phase space, it is among the
widest known binaries (Nesvorn{\'y} et al. 2003, Nicholson et al. 2008,
Noll et al 2008a).

Widely separated binaries were probably much more common in the past
as they are unstable over the age of the solar system from dynamical
perturbations (Petit and Mousis 2004; Noll et al. 2006, 2008a; Petit
et al. 2008; Parker and Kavelaars 2010).  Its possible the cold
classical KBOs have had lower numbers of collisions or fewer planet
interactions allowing the largest population of surviving wide,
equal-sized binaries in the Kuiper belt.  2007 TY430 has similar
separation and component sizes as the cold classical KBOs 1998 WW31
and 2000 CF105 (Table 5).  Petit and Mousis (2004) determined these
two binary KBOs were only stable for about 1-2 Gyrs.  The Petit and
Mousis (2004) result is likely a lower limit on the binary
lifetimes since they used a steeper size frequency distribution for
the smaller Kuiper belt objects than currently believed to exist,
allowing for more small impactors to disrupt the binaries than is
currently likely (Fraser et al. 2008; Fuentes and Holman 2008; Fuentes
et al. 2009; Fraser 2009).  The 2007 TY430 binary probably has a
similar timescale of stability, though 2007 TY430 is in the Plutino
population and not the low inclination classical belt.  The timescales
of destruction for wide binary TNOs by Parker and Kavelaars (2010)
also suggest 2007 TY430 would not have a high probability for
surviving for the age of the Solar System, and thus 2007 TY430 could
just be one of a few binaries left of a once much larger population of
binaries.

\subsubsection{2007 TY430: The Case for a Primordial Binary Orbit}

The inclinations of many TNO binary components are generally high
(Naoz et al. 2010).  The Kuiper Belt binary 2001 QW322 was found to
have a fairly low eccentricity ($e<0.4$) yet a very large inclination
($i\sim125$ degrees) and semi-major axis ($\sim 120000$ km) (Petit et
al. 2008).  Because of 2001 QW322's large inclination, its relatively
low eccentricity and large semi-major axis can be explained by the
Kozai mechanism (Perets and Naoz 2009).  Most binary minor planets do
not appear to have primordial orbits as they could have been affected
by either tidal forces or have high eccentricities or high
inclinations at which the Kozai mechanism may operate (Ragozzine 2009,
Naoz et al. 2010).

The binary orbit of 2007 TY430 is compared to other known equal-sized
binary orbits in Figures ~\ref{fig:binarymutualrhi} and
~\ref{fig:binarymutualrhe}.  2007 TY430 does not have orbital elements
similar to other TNO binaries shown in a similar Figure 3 of Naoz et
al. (2010), which compares the normalized separations of the binary
components to their eccentricities.  The $log(a_{bin}/r)$ for 2007
TY430 is slightly greater than 2.5, but still in a stable region of
the Naoz et al. Figure 3.  Interestingly, when including recent
results from Parker et al. (2011), it appears most of the equal-sized
binaries with very large semi-major axes appear to have moderate to
low eccentricities and low inclinations, making them unsusceptible to
the Kozai resonance and tides.  This is in contrast to many of the
lower semi-major axis, equal-sized binaries that appear susceptible to
the Kozai mechanism because of their large mutual inclinations and
higher eccentricities.  This result is likely because the large
semi-major axis objects would become unstable to perturbations if
their eccentricities or inclinations were too large.

2007 TY430 is the lowest eccentricity, wide component, equal-sized
binary known.  Because the mutual orbital elements of the 2007 TY430
binary has low inclination, low eccentricity and large semi-major axis
(Figures~\ref{fig:binarymutualrhi} and ~\ref{fig:binarymutualrhe}),
the Kozai mechanism and tidal interactions have not likely modified
the primordial orbit of 2007 TY430 (Chyba et al. 1989; Murray and
Dermott 1999; Perets and Naoz 2009).  Modification of the 2007 TY430
binary could have occurred over the age of the solar system from
direct collisions on either of the two components, from relatively
massive bodies passing within the Hill sphere of the two components or
interactions with the giant planets (Petit and Mousis 2004;
Nesvorn{\'y} et al. 2011).

The high eccentricities and inclinations of other TNO binary
components suggest formation in a dense collisional environment.  In
this environment, gravitational encounters can create these high
eccentricities and inclinations along with secular Kozai effects and
tidal evolution (Naoz et al. 2010).  Other binaries observed to date
could have formed in a similar manner as 2007 TY430, but later three
body encounters or direct collisions after binary formation changed
the binary components orbits to be highly inclined and eccentric.  It
is thus possible that 2007 TY430 simply has escaped any significant
collisions or gravitational encounters while other known binaries have
not or formed in a different collisional environment.  This could be
because 2007 TY430 diffused out of the low inclination classical belt
early on and the Plutino orbit made it less susceptible to collisions.
Thus 2007 TY430's binary orbit could be more primordial than many of
the other known equal-sized wide binary objects observed to date.

\subsection{Formation of the 2007 TY430 binary system}

The proposed binary formation by Funato et al. (2004) from exchanged
binaries is unlikely since TNOs appear to lack the many
high-eccentricity binaries suggested by this mechanism (Naoz et
al. 2010), of which 2007 TY430 is another example of a low
eccentricity binary.  Weidenschilling (2002) proposed a hybrid
mechanism of a direct collision between two bodies while in the Hill
Sphere of a third body, which can directly form large separation
binaries.  This mechanism would only be likely to occur during the
formation of the Kuiper Belt, as many more large bodies than currently
observed are required as well as low velocities are needed.  A similar
mechanism, the L$^{3}$ mechanism proposed by Goldreich et al. (2002),
has a third body strongly interacting with two other bodies while they
each are within the other body's Hill sphere.  The L$^{3}$ mechanism
should dominate over the Weidenschilling (2002) mechanism because
strong gravitational interactions should occur much more frequently
than the actual collisions of objects.  Chaos assisted capture
(Astakhov et al. 2005, Lee et al. 2007), is effectively the same as
the L$^{3}$ mechanism; in chaos assisted capture, two bodies are
temporarily trapped in their mutual Hill Spheres and can become
permanently trapped when a third 'intruder' body is scattered by the
pair.  This mechanism would create wide separation binaries with
equal-size and moderate eccentricity.

A second binary formation mechanism proposed by Goldreich et
al. (2002), the L$^{2}$s mechanism, involves two objects forming an
unbound transient binary that becomes bound with the aid of the
dynamical friction from a sea of small objects.  Any binary formation
that involves dissipation of energy in a smooth and gradual manner,
like the L$^{2}$s mechanism, will likely form retrograde binaries
(Schlichting and Sari 2008b).

Because of 2007 TY430's prograde orbit, it is unlikely to have formed
by the L$^{2}$s mechanism.  Unlike the L$^{2}$s mechanism, the L$^{3}$
mechanism of Goldreich et al. (2002) should form equal populations of
prograde and retrograde binaries (Schlichting and Sari 2008).  So 2007
TY430 could have formed through the L$^{3}$ mechanism which would only
be likely if the Kuiper Belt objects had relatively large velocities,
called super-Hill velocities (Schlichting and Sari 2008a, 2008b).

The Hill velocity (Rafikov 2003; Goldreich et al 2004; Murray-Clay and
Chiang 2006; Lee et al. 2007) is defined as:

\begin{equation}
v_H = \left[\frac{G(m_{p1}+m_{p2})}{r_{H}}\right]^{1/2}
\label{eq:hill}
\end{equation}

where $G$ is the gravitational constant.  For 2007 TY430, the Hill
velocity is about 0.4 m s$^{-1}$ which is much less than the Keplerian
velocity of its Plutino type orbit that is about 4 km s$^{-1}$.

Formation through the L$^{3}$ mechanism would require large velocities
between KBOs that several authors do not think were prevalent in the
early Kuiper Belt (Goldreich et al. (2002), Schlichting and Sari
(2008b), Murray-Clay and Schlichting (2011)).  Goldreich et al. (2002)
suggest the velocities of KBOs were about a third of the Hill
velocity, which would make retrograde orbits through the L$^{2}$s
mechanism the dominant binary formation mechanism.  This is because
binary formation efficiency from gravitational encounters decreases
significantly with higher relative velocities since the sphere of
influence or time an object affects another is decreased (Noll et
al. 2008a; Schlichting and Sari 2008b).  Murray-Clay and Schlichting
(2011) argue that sub-Hill velocities and the binary formation from
dynamical friction (L$^{2}$s mechanism) is the best mechanism for the
equal-sized binary formations.  2007 TY430 does not fall in along
these formation lines because it is not retrograde, meaning large
velocities during formation were likely and thus dynamical friction
unlikely to be the capture mechanism.  A weakness to the high velocity
scenario is that Schlicting and Sari (2008b) suggest that the velocity
would need to be finely tuned and stay in such a state for a
considerable amount of time in order for a high velocity regime to
form these prograde, equal-sized binaries.

To date, not enough equal-sized binary orbital inclinations are known
to draw a strong conclusion.  Only 17 equal-sized KBO binaries have
known inclinations (Figure~\ref{fig:binarymutualrhi}).  Of these five
are retrograde and twelve are prograde, but three of the prograde
objects have $\Delta Mag>1$ mag and thus are not quite equal-sized
binaries.  Several authors have suggested that the observed
inclinations of TNO binaries are consistent with them being randomly
distributed (Noll 2003, Chiang et al. 2006, Grundy et al. 2011).  This
suggests the Kuiper Belt disk was in the dispersion-dominated
(dynamically hot) regime during binary formation (see Stewart and Ida
2000; Collins and Sari 2006; Schlichting and Sari 2008b).  Lee et
al. (2007) point out there are more prograde than retrograde binaries
and determine this as a sign of how the binaries formed.  Schlichting
and Sari (2008b) suggest this sign is showing the velocity regime in
which the binaries formed.  Schlichting and Sari (2008b) predicted
that over 97 percent of binaries with comparable masses will have
retrograde orbits if the relative velocities of the Kuiper Belt
objects were low.  Since it appears to be well below 50 percent, the
relative velocities must have been significantly higher, as discussed
above for the formation of 2007 TY430.

Binary formation from direct gravitational collapse of an over density
of concentrated cm to meter sized solids was suggested by Nesvorn{\'y} et
al. (2010).  This mechanism appears to be able to produce a wide range
of distant equal-sized binaries with a large range of eccentricities
with most having prograde inclinations.  This mechanism could explain
2007 TY430, but may have trouble producing the retrograde binaries
found to date.

\subsection{Plutino Resonance and the Cold Classical Belt}

The Neptune mean motion resonance populations could have been emplaced
by one of two favored mechanisms.  1) The slow migration outwards of
Neptune by over 10 AU from its formation location from planetesimal
scattering would have allowed Neptune to sweep many objects into the
resonances we see today (Malholtra 1995; Hahn and Malhotra 2005).
This scenario would likely mean that the inner 3:2 resonance would
have a significantly different population of objects than the more
outer resonances that would have swept through the classical Kuiper
Belt population.  This scenario would also suggest the resonance
populations should have a cold component representative of the cold
classicals along with a hot component representing the objects from
closer in.  2) The chaotic population of the resonances could have
occurred if Neptune was scattered with a relatively large eccentricity
out from near the current Saturn region to near its present semi-major
axis (Levison and Morbidelli 2003; Tsiganis et al. 2005; Levison et
al. 2008).  As Neptune circularized its orbit through interactions
with planetesimals, objects in the resonance areas would become
trapped.  In this scenario, the resonance populations would be more
uniform and not have a cold component, unlike the slow migration
scenario.

The resonance populations should have a cold component if Neptune
experienced extensive slow, smooth migration.  This cold component
should be similar in characteristics to the low inclination classical
Kuiper Belt, that is, equal-sized binaries should be common as well as
ultra-red material (Murray-Clay and Schlichting 2011).  This cold
component should also have lower eccentricities compared to other
objects, meaning there could be a correlation between binaries, color
and eccentricity as well as with inclination in the resonance
populations.  In the Levison et al. (2008) scenario, Chaotic
scattering of Neptune and then circularization, the resonance
populations should match the scattered disk in characteristics.

No obvious cold component is observed in the 3:2 resonance population
and the statistics are too low for the other resonance populations to
determine if there is a cold component (Murray-Clay and Schlichting
2011).  The 3:2 having no cold component is still consistent with
slow, smooth migration since the 3:2 does not sweep through the cold
classicals (Murray-Clay and Schlichting 2011).  Thus the 3:2 resonance
is not as strong as a marker as the other resonances, which should
have cold components as they swept through the classical region of the
Kuiper Belt.  It is possible that sweeping of the $\nu_{8}$ resonance
through the 3:2 resonance and inner classical Kuiper belt cleared out
much of these regions (Petit et al. 2011).  If this is true, the
$\nu_{8}$ resonance likely did not sweep through until after the
formation of any equal-sized binaries since the binaries are unlikely
to have formed as the density of such objects would have been too low.

2007 TY430 has a moderately inclined and eccentric heliocentric orbit,
and thus would be considered part of the hot component of the 3:2
resonance on dynamical grounds. Although 2007 TY430 could easily have
been emplaced recently by dynamical scattering and resonance sticking,
it is interesting to speculate on the meaning of this cold-classical
like binary if it was placed in the resonance primordially. As
discussed above, if 2007 TY430 was captured from an originally low
eccentricity, low inclination orbit by slow, smooth migration, it
would have originated around 31 AU by conservation of Brouwer's
integral. This would suggest that the primoridal cold classical region
extended much further inward than is seen today, but the observational
evidence for a cold region of the 3:2 resonance is insufficient
(Murray-Clay and Schlichting 2011).  If its dynamical history was
well-known, the relatively high eccentricity and inclination of 2007
TY430 could have been seen as an indication that the slow, smooth
migration model is insufficient.  More binary objects need to be
discovered in the 3:2, but the existence of 2007 TY430 in the hot
region of the 3:2 and no other known equal-sized 3:2 resonance
objects, would favor chaotic scattering by Neptune over slow smooth
migration.

\subsection{Other Known Equal-Sized Wide Binaries}

\begin{figure}
\epsscale{0.4}
\centerline{\includegraphics[angle=90,totalheight=0.6\textheight]{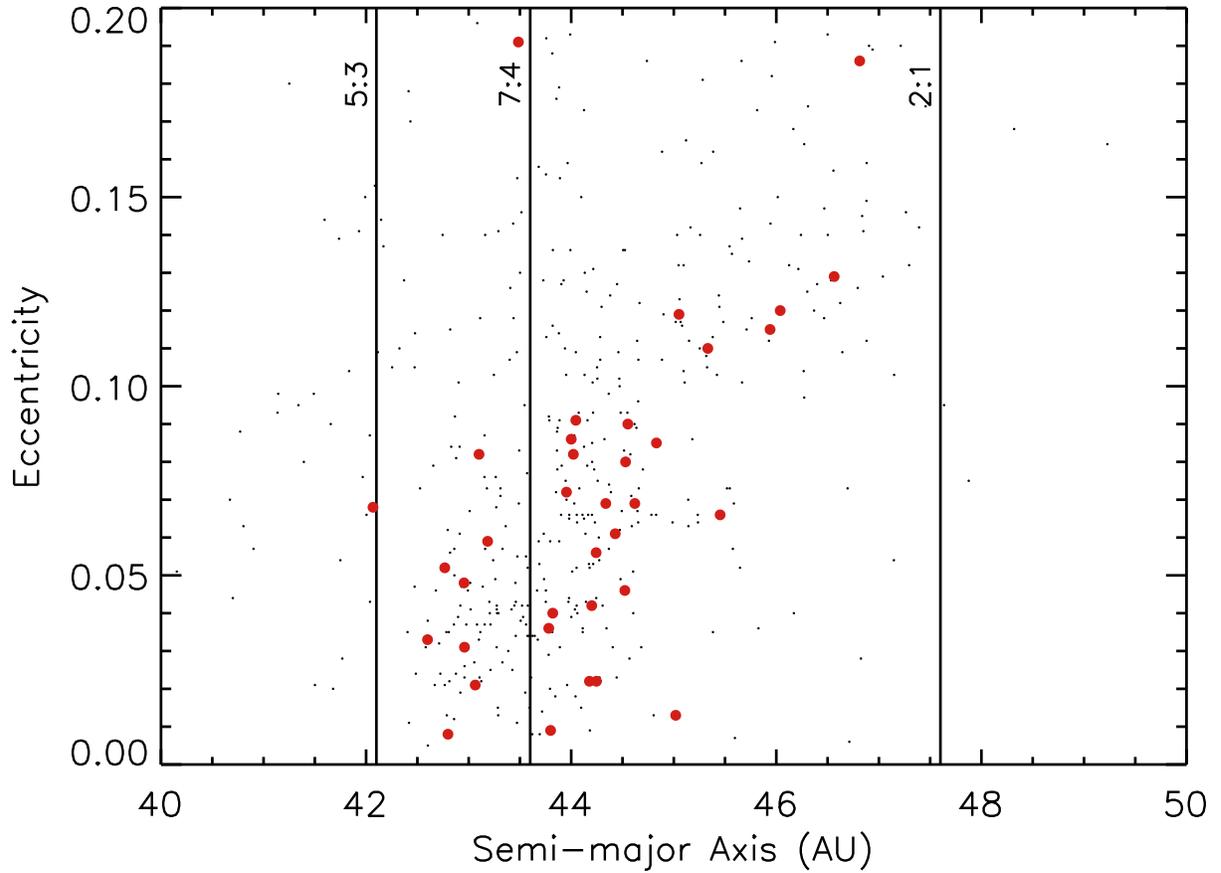}}
\caption{An expanded version of Figure~\ref{fig:kboeabinary} to better
  see the distribution of equal-sized binaries in the main classical
  Kuiper belt.  No enhancement of binaries is seen near 44 AU, which
  has been called the kernel area by Petit et al. (2011).  There is an
  apparent lack of binaries near 43.5 AU, which is near the 7:4
  resonance as well as above an eccentricity of 0.1.}
\label{fig:kboeabinaryblowup} 
\end{figure}

\begin{figure}
\epsscale{0.4}
\centerline{\includegraphics[angle=90,totalheight=0.6\textheight]{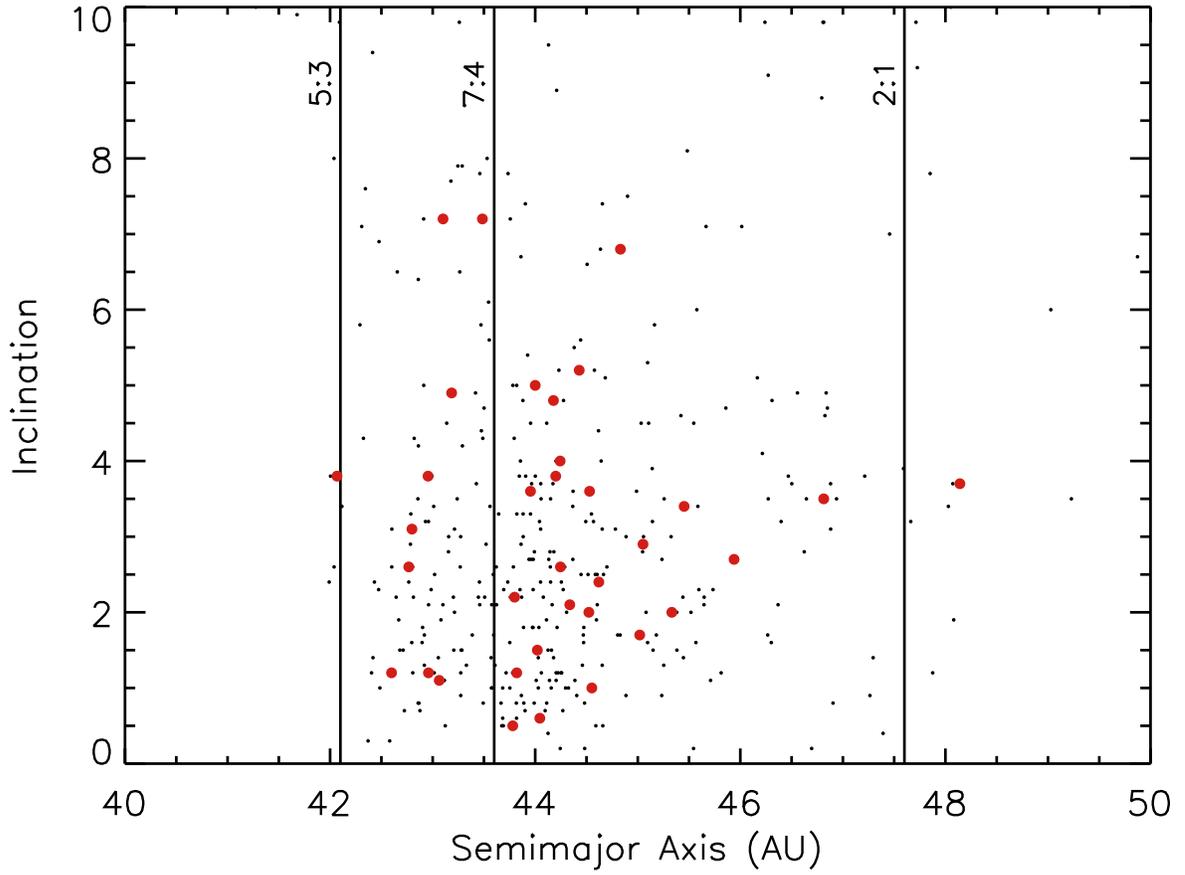}}
\caption{A blown-up version of Figure~\ref{fig:kboiabinary} to better
  see the distribution of equal-sized binaries in the main classical
  Kuiper belt.  Most binaries have inclinations less than about 5 degrees.}
\label{fig:kboiabinaryblowup} 
\end{figure}

\begin{figure}
\epsscale{0.4}
\centerline{\includegraphics[angle=90,totalheight=0.6\textheight]{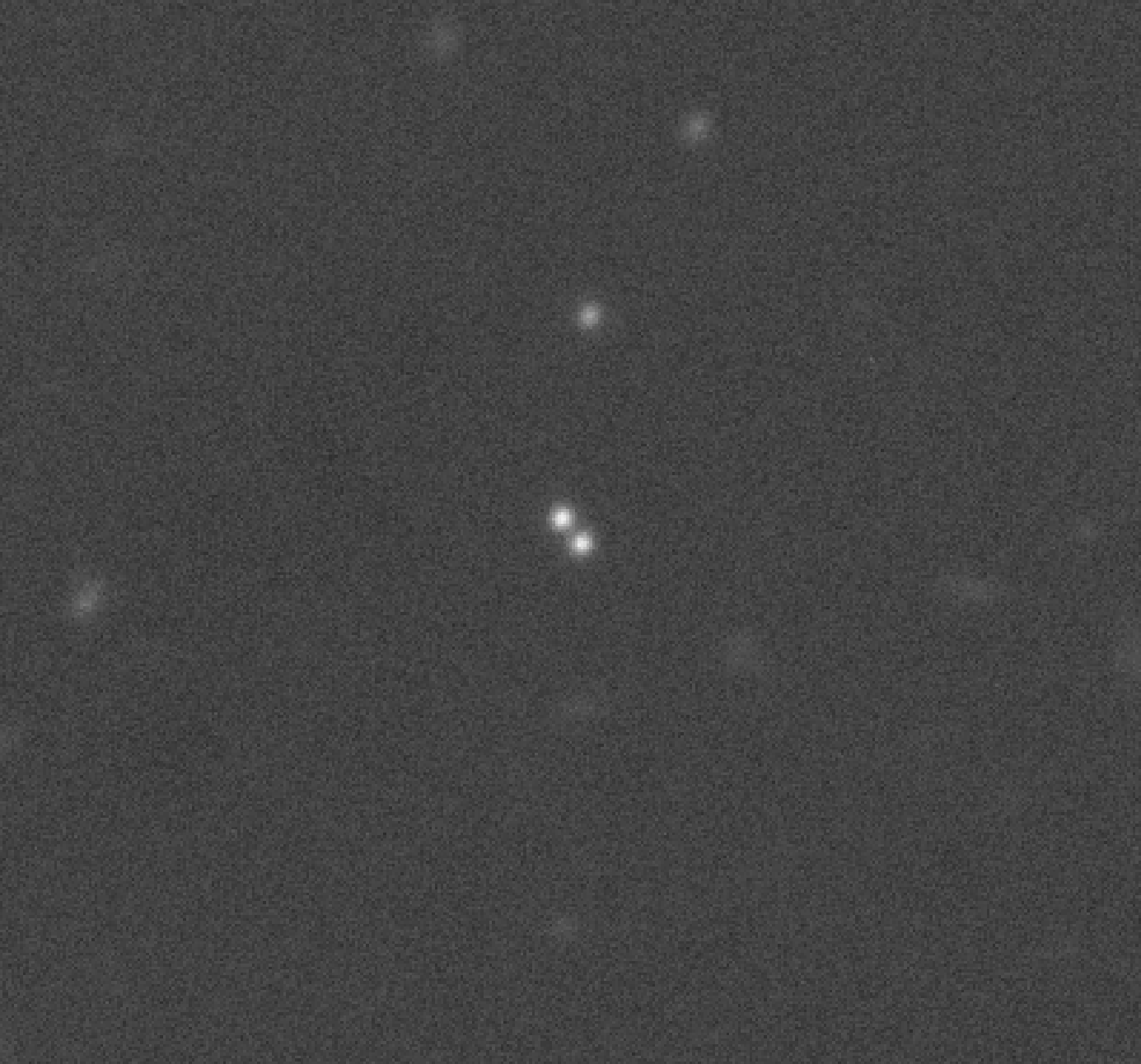}}
\caption{The equal-sized Plutino binary 2007 TY430 is easily seen near
  the center of this image from the GMOS detector on the Gemini
  telescope in December 2007.  The separation between the components
  was about 0.7 arcseconds.}
\label{fig:image2007TY430} 
\end{figure}

\begin{figure}
\epsscale{0.4}
\centerline{\includegraphics[angle=0,totalheight=0.6\textheight]{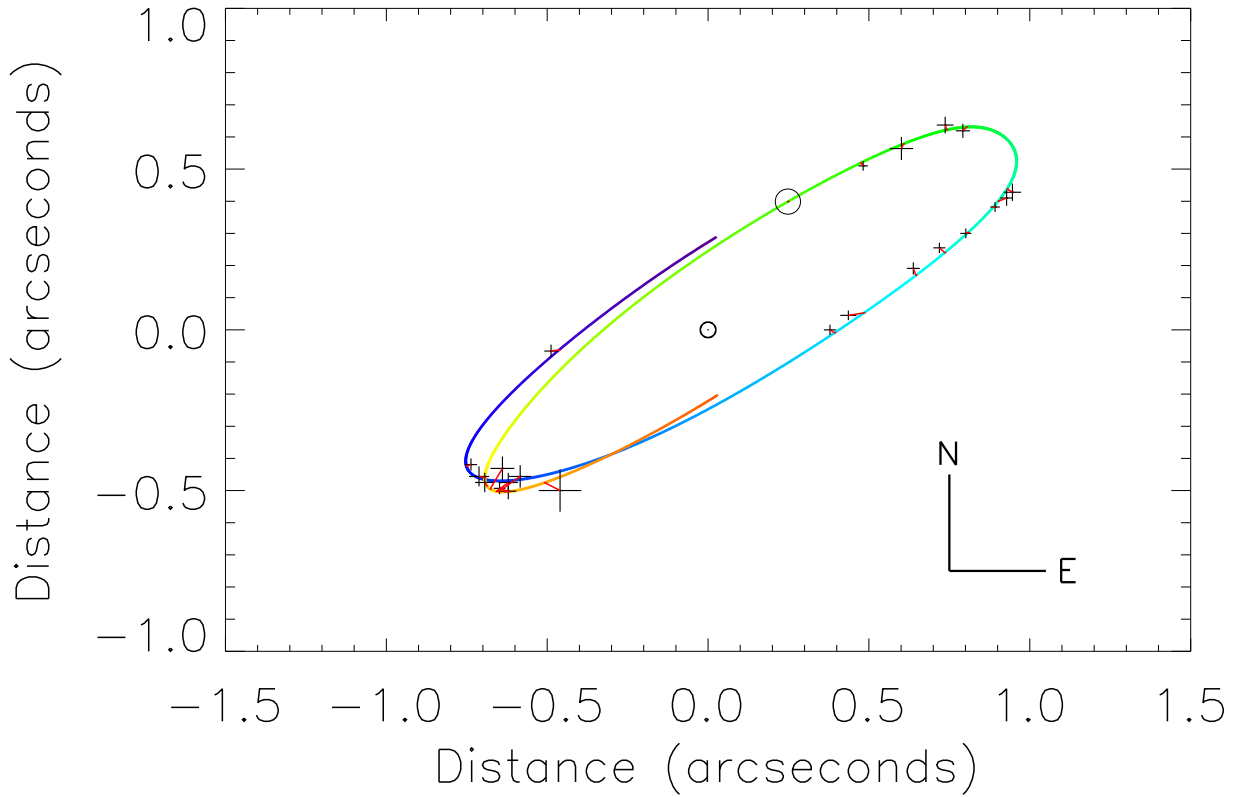}}
\caption{The orbital fit using the observations of 2007 TY430 from the
  Gemini telescope and HST between November 2007 and January
  2011. This plot shows the relative motion of one component around
  the other (which is held fixed at the center). Small crosses show
  the detected resolved component astrometric observations from Table
  2 and are connected by red lines to the predicted position of these
  observations from the best-fit orbital parameters (Table 3) which
  are strongly constrained by the high precision astrometry and the
  excellent phase coverage. A line traces out the relative orbit over
  the timespan of observations with color representing time (red
  starting 100 days before the first observation and purple ending 100
  days after the last observation). Parallax and proper motion cause
  the apparent orbit to deviate from a closed Keplerian ellipse.
  Because of the very small error bars on the HST observation, it is
  highlighted with a circle.  The dot at the center is a to scale
  component of 2007 TY430 (50 km radius or 2.4 mas) with a circle ten
  times the normal size to show its location.}
\label{fig:orbitfit} 
\end{figure}

\begin{figure}
\epsscale{0.4}
\centerline{\includegraphics[angle=0,totalheight=0.6\textheight]{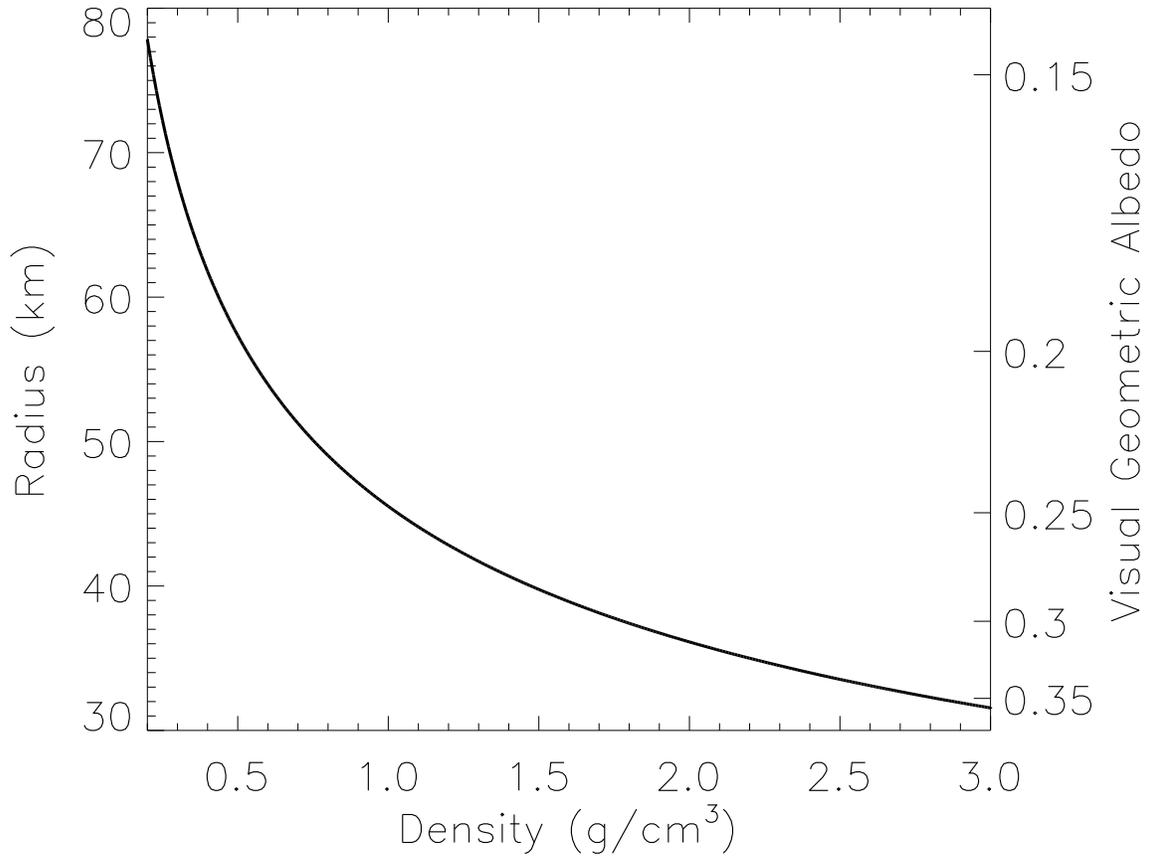}}
\caption{The density versus radius (left y-axis) and visual geometric
  albedo (right y-axis) for each of the (assumed identical) components
  of 2007 TY430.  In order to have a reasonable density ($\rho > 0.5$
  g/cm$^{3}$) the albedo is required to be at least moderate
  ($p_{v}>0.17$).  This minimum density also requires the two components
  to have radii less than 60 km.}
\label{fig:densityradius} 
\end{figure}

\begin{figure}
\epsscale{0.4}
\centerline{\includegraphics[angle=90,totalheight=0.6\textheight]{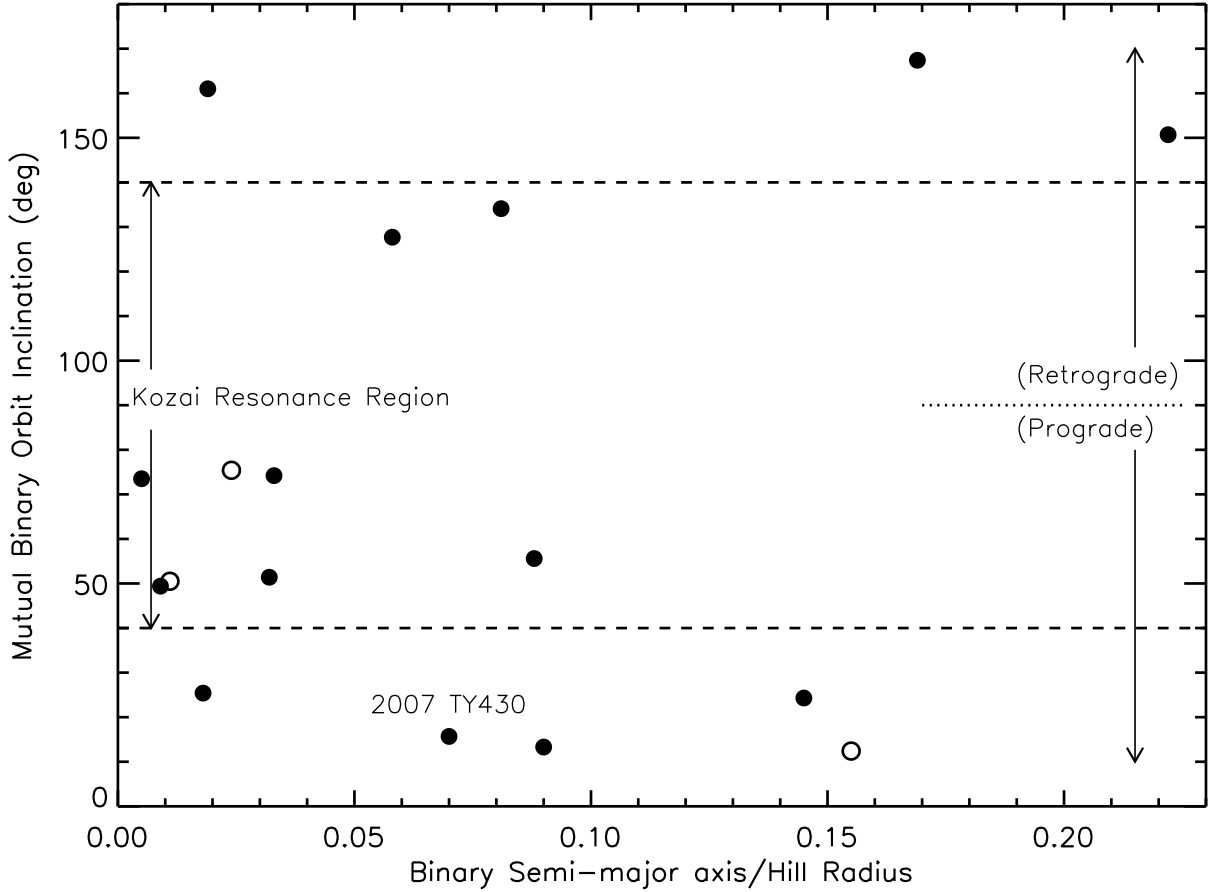}}
\caption{The semi-major axis/Hill radius ($a_{bin}/r_{H}$) versus the
  mutual orbit inclination with respect to the J2000 ecliptic for the
  known KBO equal-sized binaries with known mutual inclinations.  KBOs
  (26308) 1998 SM165, (42355) Typhon and 2005 EO304 are shown with
  open symbols because they are not technically equal-sized binaries
  as their $\Delta$Mag are greater than 1 mag, but they all have large
  semi-major axes and so are included here.  Many of the KBO binaries
  have likely been influenced by the Kozai mechanism.  2007 TY430 is
  in the lower center of this figure.  The Kozai resonance likely
  affects distant circular orbits between about 40 and 140 degrees
  (dashed lines) based on irregular satellites and numerical
  simulations (Kozai 1962; Carruba et al. 2002; Nesvorn{\'y} et
  al. 2003; Sheppard et al. 2006).  Interestingly, most of the very
  large semi-major axis, equal-sized binaries appear to have moderate
  to low eccentricities and low inclinations, making them
  unsusceptible to the Kozai resonance and tides.}
\label{fig:binarymutualrhi} 
\end{figure}

\begin{figure}
\epsscale{0.4}
\centerline{\includegraphics[angle=90,totalheight=0.6\textheight]{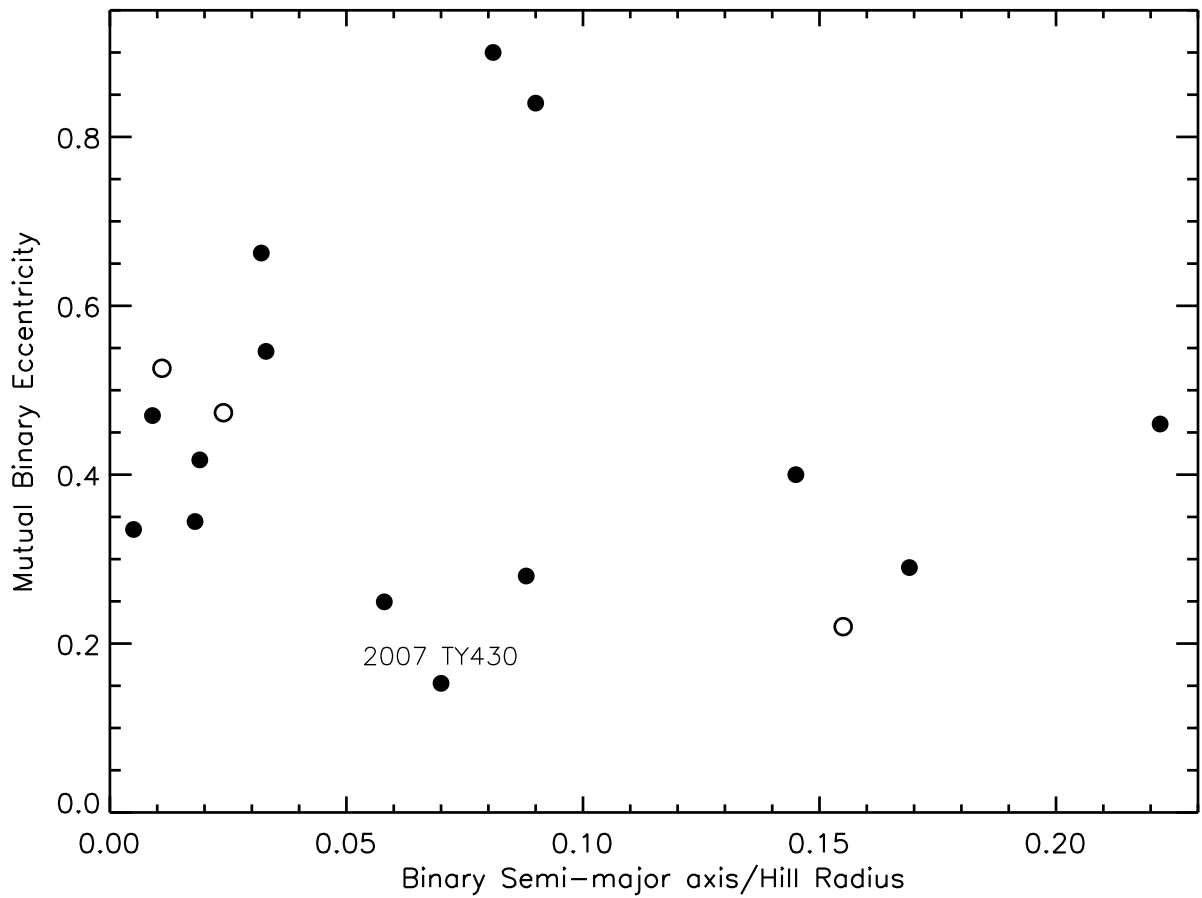}}
\caption{The semi-major axis/Hill radius ($a_{bin}/r_{H}$) versus the
  mutual orbit eccentricity for the known KBO equal-sized binaries
  with known mutual inclinations.  2007 TY430 in the lower middle
  portion and has the lowest eccentricity of any known equal-sized
  binary, this along with 2007 TY430's low prograde inclination make
  it unlikely that tides or the Kozai resonance have significantly
  altered its binary orbit over the age of the solar system.  Again, KBOs
  (26308) 1998 SM165, (42355) Typhon and 2005 EO304 are shown with
  open symbols because they are not technically equal-sized binaries
  as their $\Delta$Mag are greater than 1 mag, but they all have large
  semi-major axes and so are included here.}
\label{fig:binarymutualrhe} 
\end{figure}

Table 5 shows the other known equal-sized TNO binaries.  Here,
equal-sized is defined as the components having less than 1 magnitude
difference between components.  This magnitude difference corresponds
to mass ratios less than 4, assuming similar albedos and densities for
the two components and the lack of significant rotational brightness
variations.  As shown in Table 5 and Figures~\ref{fig:kboeabinary}
to~\ref{fig:kboiabinaryblowup}, the equal-sized binary population is mostly
in the low inclination classical belt. Interestingly, most of the
equal-sized binary TNOs not in the low inclination classical belt
appear to be in mean motion resonances.  We see no obvious increase in
the binary population between the semi-major axes of 43.5 and 44.5 AU,
which has been called the ``Kernel'' area by Petit et al. (2011) as
this region appears to be somewhat more dense than the rest of the
classical Kuiper belt (Figures~\ref{fig:kboeabinaryblowup}
and~\ref{fig:kboiabinaryblowup}).  It does appear that there is an
absence of equal-sized binaries around 43.5 AU, which is near the 7:4
resonance, as well as very few with eccentricities greater than 0.10.

\begin{figure}
\epsscale{0.4}
\centerline{\includegraphics[angle=90,totalheight=0.6\textheight]{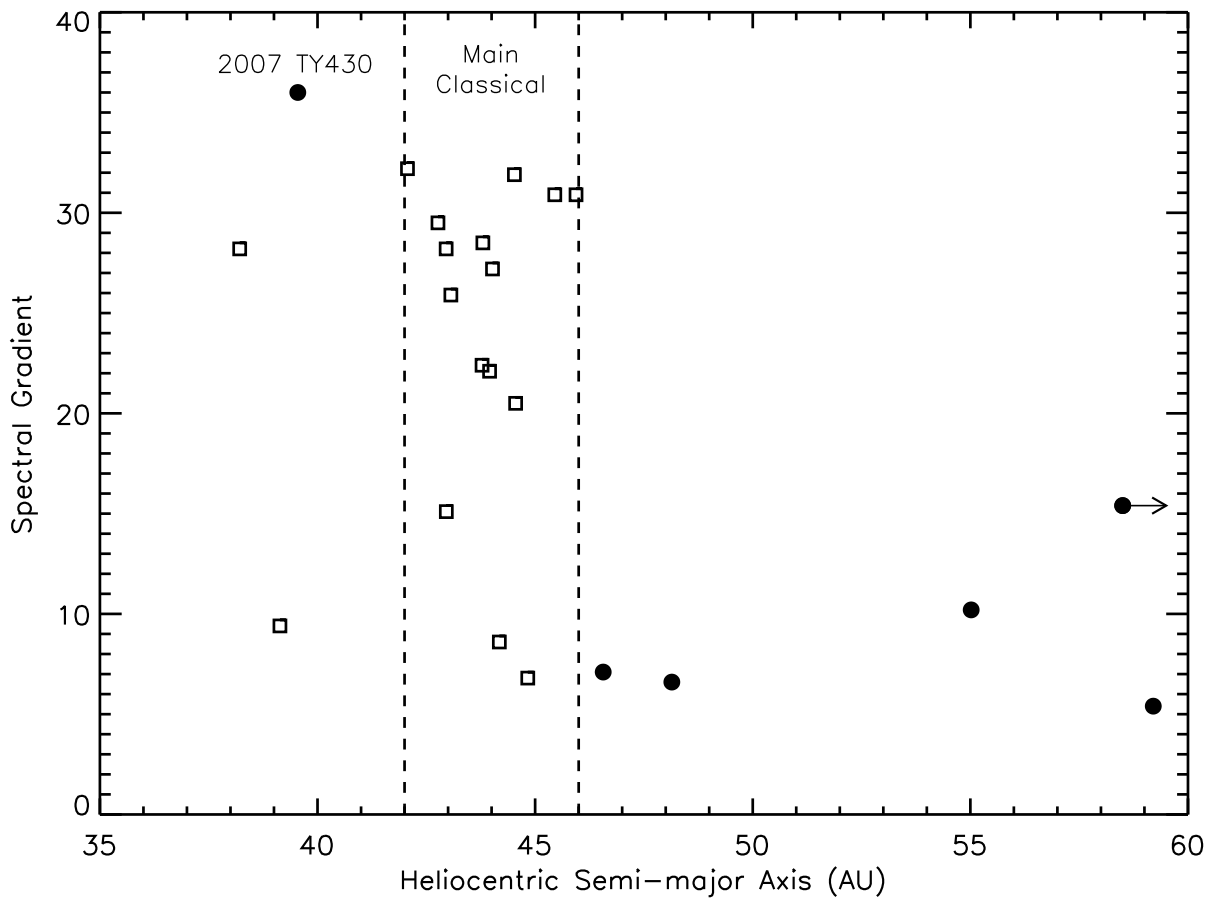}}
\caption{The spectra gradients of the equal-sized binaries with known
  colors is shown versus the heliocentric semi-major axis.  The
  classical Kuiper Belt objects are shown with squares while the
  resonant and scattered disk objects are shown with filled circles.
  Most of the main classical Kuiper Belt objects between about 42 and
  46 AU have very red or ultra-red colors while only 2007 TY430 shows
  an ultra-red color outside of the classical belt.  Both 1998 WV24
  and 1999 OJ4 are a little closer in AU than the main classical belt,
  but since both have relatively low eccentricities they are consider
  part of the inner classical belt.  Thus the inner classical Kuiper
  belt is likely an extension of the main classical Kuiper belt as
  both show ultra-red, equal-sized binaries.  Since all of the
  equal-sized binaries outside the classical belt, except for 2007
  TY430, do not have ultra-red colors, it likely means that
  equal-sized binaries formed in multiple locations and not just in
  the cold classical belt.}
\label{fig:binarycolors} 
\end{figure}

The colors of the equal-sized binaries are not all ultra-red, as many
are near neutral in color (Tegler and Romanishin 2000; Trujillo and
Brown 2002; Gulbis et al. 2006; Peixinho et al. 2008; Petit et
al. 2008; Benecchi et al. 2009; Sheppard 2010).  As
Figure~\ref{fig:binarycolors} shows, the only equal-sized binary that
is ultra-red in color ($S>25$ as defined in Sheppard (2010)) and not
in the classical population is 2007 TY430 (colors are from the
database described in Hainaut and Delsanti 2002).  In all, only 1 of 6
equal-sized binaries with known colors outside of the classical
population has an ultra-red color, which is 2007 TY430 (Table 5).  In
contrast 13 of 17 equal-sized binaries in the classical population
have spectral gradients near or within the ultra-red color region
($S>20$).

Numerically integrating the non classical equal-sized binaries from
Table 5 (2000 QL251, 2000 FE8, 2006 SF369, 1998 WV24, 2000 CM114, 2001
QC298) with ten clones each for 500 MYr found all but 2000 CM114 were
fairly stable. Some of the clones of 2000 QL151 ($6/10$, 1 a little
chaotic) and 2006 SF369 ($10/10$, 3 a little chaotic) were
experiencing Kozai oscillations (not unusual for resonant objects).
Thus only 2000 CM114 is likely to have come from the classical region,
but its inclination is currently very large and it is not ultra-red in
color. Since all of the equal-sized binaries outside the classical
belt, except for 2007 TY430, do not have ultra-red colors, it suggests
that equal-sized binaries formed in multiple locations and not just in
the cold classical belt.  As shown in Figure~\ref{fig:binarycolors},
the inner classical Kuiper belt is likely an extension of the main
classical Kuiper belt as both have low inclination, ultra-red,
equal-sized binaries.  If the ultra-red, equal-sized binary of 1999
OJ4 formed in the inner classical belt, where we see it today, it must
have formed before any clearing through the $\nu_{8}$ resonance
sweeping mechanism since the desnity of KBOs would have been to low
for likely equal-sized binary formation after any significant clearing
of the region.

%Figure~\ref{fig:binarymutualei}

%??? How many of the known wide binaries (including those in/near the
%cold classicals) are in the "kernel" that the CFEPS team has? The
%kernel is at a,e,i of about 44.2, 0.07, 2 degrees. I believe them when
%they say that it is not just a smooth extension of the cold classical
%population, but I haven't seen anyone do any analysis on differences
%between the kernel and cold classicals either in color or in
%binarity.

%(42355) Typhon/Echidna has a prograde orbit (Grundy et al. 2008).
%(134860) 2000 OJ67 and 2004 PB108 have nearly polar orbits (Grundy et
%al. 2009). 2001 QW322 has a retrograde orbit (Petit et al. 2008).

% Mean Longitude ($\lambda \equiv \Omega + \omega + M = 224.9\pm 1.1 $ deg.

%The Mean Longitude is not critical; the main purpose in included it is
%to show (indirectly) that the error in the argument of periapse and
%mean anomaly are highly correlated, such that the combination in the
%form of mean longitude has much smaller error. It can be included in
%the text or in a table, but I see no harm in putting it in.

\section{Summary}

2007 TY430 is the first known wide, equal-sized binary with an
ultra-red color outside of the classical Kuiper belt.  The binary is
near the outer edge of the 3:2 mean motion resonance with Neptune and
could have become ``stuck'' at its current location after escaping
from the main classical Kuiper Belt.

1) The binary components of 2007 TY430 were observed on approximately
a monthly basis between 2007 and 2011 with the Gemini telescope in
excellent seeing conditions.  The components had a maximum separation
of about 1 arcsecond with an average of about 0.1 magnitude difference
between them.

2) Through the extensive observations, the mutual orbit of 2007
TY430's components were found to have a large semi-major axis
($21000\pm160$ km), moderate to low eccentricity ($0.1529\pm0.0028$)
and low prograde inclination ($15.68\pm0.11$ deg) with the mirror
orbit rejected with high confidence.  The low inclination, large
semi-major axis and relatively low eccentricity of the mutual binary
components means neither tides nor the Kozai mechanism should have
significantly altered the binary orbit of 2007 TY430.  The possible
primordial orbit for the components of 2007 TY430 is unique compared
to many other known binary KBOs.  It appears that the KBO binaries
with the largest mutual component semi-major axes have on average
lower eccentricities and inclinations compared to the smaller
semi-major axis binaries.  This is likely because the largest
semi-major axis objects would become unstable to perturbations if
their eccentricities or inclinations were too large.

3) The favored formation mechanism for the low inclination prograde
orbit of 2007 TY430 is the L$^{3}$ mechanism proposed by Goldreich et
al. (2002).  Under the model of Schlichting and Sari (2008), this
would suggest the Kuiper Belt objects had relatively large velocities
relative to each other during binary formation.  The gravitational
collapse mechanism of binary formation proposed by Nesvorny et
al. (2010) is also a good fit for the prograde binary 2007 TY430, but
this mechanism has not yet been shown to work similarly on the known
retrograde equal-sized binaries.

4) The wide binary nature of 2007 TY430, when compared to previous
simulations of similarly separated objects, suggests 2007 TY430 is
likely not a stable binary for the age of the solar system.  Thus 2007
TY430 could just be one of the few remaining binaries from a once much
larger population.  Since all of the equal-sized binaries outside the
classical belt, except for 2007 TY430, do not have ultra-red colors,
it likely means that equal-sized binaries formed in multiple locations
and not just in the cold classical belt.  The inner classical Kuiper
belt is likely an extension of the main classical Kuiper belt as both
have low inclination, ultra-red, equal-sized binaries.  If the
ultra-red, equal-sized binary of 1999 OJ4 formed in the inner
classical belt, where we see it today, it must have formed before any
clearing from any possible $\nu_{8}$ resonance sweeping.

5) The total system mass of $7.90\pm0.21 \times 10^{17}$ kg means 2007
TY430 has a moderate to high albedo and low density.  Assuming a
reasonable density of 0.75 g/cm$^{3}$ results in an albedo of about
0.23 with radii of the individual components at about 50 km.  The
individual components of 2007 TY430 were measured to have identical
ultra-red colors.  No obvious water or methane ice signatures were
detected on the surface of 2007 TY430 from infrared observations using
special water and methane ice near-infrared filters.

6) Out of about 1000 faint KBOs detected in ultra-deep surveys with
the Subaru and Magellan telescopes, only 2007 TY430 was found to be
an obvious wide binary.  2007 TY430 was the brightest object detected
of the 1000 objects.  Combining this result with previous works
suggests that the larger KBOs are more likely to have wide binaries,
but further observations of the faintest objects with HST is required
to confirm this.

%This would suggest that 2007 TY430's formation environment is
%likely the cause of its ultra-red color and wide binary nature and not
%its current orbital characteristics.

%under the model of Schlichting and XXX (20XX), the low inclination of
%2007 TY430 suggests super-Hill velocities for KBOs during binary
%formation.

\section*{Acknowledgments}
Based in part on data collected at Subaru Telescope, which is operated
by the National Astronomical Observatory of Japan.  Based in part on
observations obtained at the Gemini Observatory that supported C.T.,
and is operated by the Association of Universities for Research in
Astronomy, Inc., on behalf of the international Gemini partnership of
Argentina, Australia, Brazil, Canada, Chile, the United Kingdom, and
the United States of America. Data were collected under Gemini
programs GS-2007B-DD-4, GN-2008B-Q-34 and GN-2009B-Q-25.  S. S. was
partially supported by the National Aeronautics and Space
Administration through the NASA Astrobiology Institute (NAI) under
Cooperative Agreement No. NNA04CC09A issued to the Carnegie
Institution of Washington.  S. S. was also supported in part by funds
from the NASA New Horizons Spacecraft Pluto mission.  D. R.  thanks
Matt Holman and Christian Clanton for support.

\newpage

%\documentstyle [aj_pt4]{article}    % Specifies the document style.

%\begin{document}

\begin{center}
\begin{deluxetable}{lccccccc}
%\small
\tablenum{1}
\tablewidth{5 in}
\tablecaption{2007 TY430 Heliocentric Orbital Elements}
\tablecolumns{8}
\tablehead{
\colhead{$a$}  & \colhead{$e$} & \colhead{$i$} & \colhead{$MA$} & \colhead{$\omega$} & \colhead{$\Omega$} & \colhead{Epoch} \\ \colhead{(AU)} & \colhead{} & \colhead{(deg)} & \colhead{(deg)} & \colhead{(deg)} & \colhead{(deg)} & \colhead{yyyy/mm/dd} }  
\startdata
39.562 & 0.271 & 11.3 & 352.0 & 205.0 & 196.7 & 2011/08/27\nl  %\tablenotemark{a}
\enddata
\tablecomments{
The orbital elements are from the Minor Planet Center and are the
semimajor axis ($a$), inclination ($i$), eccentricity ($e$), mean anomaly ($MA$), argument of perihelion ($\omega$), longitude of the ascending node ($\Omega$) and epoch.}
%\tablenotetext{a}{These objects could be labeled as dwarf planets since their radii are likely larger than 200 km assuming a moderate or lower albedo.}
\end{deluxetable}
\end{center}

%\end{document}             % End of document.

%
\newpage

%\documentstyle [aj_pt4]{article}    % Specifies the document style.

%\begin{document}

\begin{center}
\begin{deluxetable}{lccc}
%\small
\tablenum{2}
\tablewidth{6 in}
\tablecaption{Observations of 2007 TY430 binary components}
\tablecolumns{4}
\tablehead{
\colhead{UT Date}  & \colhead{$\Delta$ RA} & \colhead{$\Delta$Dec} & \colhead{$\Delta$Mag} \\ \colhead{} & \colhead{($\arcsec$)} & \colhead{($\arcsec$)} & \colhead{(mag)}}  
\startdata
2011 Jan 06.201  &   $ -0.488\pm0.021 $ &  $ -0.066\pm0.020$ &    $ +0.16\pm0.03$       \nl  
2010 Sep 18.533  &   $ -0.737\pm0.019 $ &  $ -0.419\pm0.019$ &    $ +0.22\pm0.01$       \nl  
2010 Feb 09.222  &   $ 0.0   \pm0.120 $ &  $  0.0  \pm0.120$ &    $...$                 \nl  
2009 Dec 24.277  &   $ 0.379 \pm0.019 $ &  $  0.000\pm0.016$ &    $ +0.08\pm0.03$       \nl  
2009 Dec 07.273  &   $ 0.436 \pm0.025 $ &  $  0.045\pm0.015$  &   $ +0.20\pm0.02$       \nl  
2009 Nov 05.353  &   $ 0.638 \pm0.020 $ &  $  0.191\pm0.019$  &   $ +0.11\pm0.02$       \nl  
2009 Oct 16.451  &   $ 0.719 \pm0.019 $ &  $  0.255\pm0.016$  &   $ +0.07\pm0.02$       \nl  
2009 Sep 27.400  &   $ 0.801 \pm0.017 $ &  $  0.300\pm0.015$ &    $ +0.07\pm0.02$       \nl  
2009 Aug 31.583  &   $ 0.892 \pm0.015 $ &  $  0.382\pm0.015$  &   $ +0.05\pm0.02$       \nl  
2009 Aug 24.622  &   $ 0.928 \pm0.020 $ &  $  0.410\pm0.024$ &    $ +0.21\pm0.02$       \nl  
2009 Aug 08.625  &   $ 0.946 \pm0.027 $ &  $  0.428\pm0.027$ &    $ +0.18\pm0.03$       \nl  
2009 Feb 01.250  &   $ 0.792 \pm0.022 $ &  $  0.619\pm0.022$ &    $ +0.25\pm0.02$       \nl  
2009 Jan 07.208  &   $ 0.737 \pm0.026 $ &  $  0.637\pm0.026$ &    $ +0.27\pm0.02$       \nl  
2008 Nov 29.191  &   $ 0.601 \pm0.036 $ &  $  0.564\pm0.036$ &    $ +0.18\pm0.03$       \nl  
2008 Oct 31.401  &   $ 0.482 \pm0.015 $ &  $  0.510\pm0.015$ &    $ +0.14\pm0.03$       \nl  
2008 Sep 20.525  &   $ 0.248 \pm0.001 $ &  $  0.399\pm0.001$  &   $ +0.22\pm0.10$       \nl  
2008 Jan 08.054  &   $ -0.712\pm0.031 $ &  $ -0.456\pm0.031$  &   $ +0.05\pm0.02$       \nl  
2007 Dec 22.031  &   $ -0.694\pm0.030 $ &  $ -0.475\pm0.030$  &   $ +0.10\pm0.03$       \nl  
2007 Dec 12.058  &   $ -0.639\pm0.037 $ &  $ -0.431\pm0.037$  &   $ +0.08\pm0.03$       \nl  
2007 Nov 30.023  &   $ -0.621\pm0.030 $ &  $ -0.475\pm0.030$  &   $ -0.06\pm0.02$       \nl  
2007 Nov 27.061  &   $ -0.584\pm0.035 $ &  $ -0.456\pm0.035$  &   $ +0.09\pm0.02$       \nl  
2007 Nov 19.181  &   $ -0.621\pm0.025 $ &  $ -0.502\pm0.025$  &   $ -0.04\pm0.02$       \nl  
2007 Nov 14.187  &   $ -0.648\pm0.018 $ &  $ -0.493\pm0.018$ &    $ -0.02\pm0.02$       \nl  
2007 Oct 14.367  &   $ -0.460\pm0.066 $ &  $ -0.500\pm0.066$ &    $...$                 \nl  
\enddata
\tablecomments{This table shows the offsets of the two components for each observation.  UT Date shows the year, month and day.  All data is from Gemini except the October 14th, 2007 discovery data from Subaru and the September 20th, 2008 observations from HST by Noll et al. (2009).  The $\Delta$Mag is the difference in magnitude between the two components.  On February 9, 2010 the individual binary components were not resolved and thus zero was chosen for the offset.}
%\tablenotetext{a}{These objects could be labeled as dwarf planets since their radii are likely larger than 200 km assuming a moderate or lower albedo.}
\end{deluxetable}
\end{center}

%\end{document}             % End of document.

\newpage

%\documentstyle [aj_pt4]{article}    % Specifies the document style.

%\begin{document}

\begin{center}
\begin{deluxetable}{ll}
%\small
\tablenum{3}
\tablewidth{3 in}
\tablecaption{Binary Orbital Elements of 2007 TY430}
\tablecolumns{2}
\tablehead{
\colhead{}  & \colhead{} }  
\startdata
$a_{bin}$     &  $21000\pm160$ km  \nl
$e_{bin}$     &  $0.1529\pm0.0028$     \nl
$i_{bin}$     &  $15.68\pm0.22$ deg  \nl
$\omega_{bin}$ &  $321.9\pm 4.0$ deg   \nl
$\Omega_{bin}$ &  $332.7\pm 1.4$ deg    \nl
$MA_{bin}$     &  $290.4\pm 4.4$  deg     \nl
$P_{bin}$      &  $961.2\pm 4.6$ days    \nl
$M_{sys}$ &  $7.90\pm0.21 \times 10^{17}$ kg  \nl
Mean Longitude  &  $224.9\pm1.1$ deg \nl
\enddata \tablecomments{ Binary orbital elements are the semimajor
  axis ($a_{bin}$), inclination with respect to the J2000 ecliptic ($i_{bin}$),
  eccentricity ($e_{bin}$), argument of periapse ($\omega_{bin}$), longitude of
  the ascending node ($\Omega_{bin}$), mean anomaly ($MA_{bin}$), period ($P_{bin}$),
  total system mass ($M_{sys}$), and Mean Longitude ($\lambda_{bin} \equiv
  \Omega_{bin} + \omega_{bin} + MA_{bin}$).  All orientation angles are given with
  respect to J2000 ecliptic.  Mean anomaly is at epoch JD 2454300.0
  (geocentric).}
%\tablenotetext{a}{These objects could be labeled as dwarf planets since their radii are likely larger than 200 km assuming a moderate or lower albedo.}
\end{deluxetable}
\end{center}

%\end{document}             % End of document.

\newpage

%\documentstyle [aj_pt4]{article}    % Specifies the document style.

%\begin{document}

\begin{center}
\begin{deluxetable}{ll}
%\small
\tablenum{4}
\tablewidth{3 in}
\tablecaption{Physical Properties of 2007 TY430}
\tablecolumns{2}
\tablehead{
\colhead{Qtty}  & \colhead{Measurement} }  
\startdata
H     &  $6.94\pm0.02$ mag combined  \nl
r     &  $21.23\pm0.01$ mag combined \nl
g-r   &  $1.07\pm0.01$ mag combined \nl
r-i   &  $0.42\pm0.01$ mag combined \nl
g-i   &  $1.49\pm0.01$ mag combined \nl
R     &  $20.94\pm0.01$ mag combined \nl 
B-R   &  $2.03\pm0.01$ mag combined \nl
V-R   &  $0.74\pm0.01$ mag combined \nl
R-I   &  $0.63\pm0.01$ mag combined \nl
B-I   &  $2.66\pm0.01$ mag combined \nl
g-i   &  $1.49\pm0.02$ mag component 1 \nl
g-i   &  $1.50\pm0.02$ mag component 2 \nl
$S$   &  $36\pm2$  combined \nl
J     &  $19.91\pm0.09$ mag combined \nl
J-H$_{2}$0 &  $-0.16\pm0.13$ mag combined \nl
J-CH$_{4}$ &  $0.07\pm0.15$ mag combined \nl
$p_{v}$  &  $\approx0.23$ assuming $\rho\sim0.75$ \nl
$r$     &  $\approx50$ km assuming $\rho\sim0.75$ \nl
\enddata \tablecomments{H is the combined absolute magnitude of the two components in the V-band.  The colors are from the Sloan g, r and i-bands and have also been converted to the Johnson-Kron-Cousins system B, V, R and I-bands using Smith et al. (2002).  $S$ is the spectral gradient of the object as described in  Sheppard (2010).  J is the J-band infrared color of 2007 TY430 and J-H2O and J-CH4 are the colors using the water and methane narrow band filters in the infrared K-band as described in Trujillo et al. (2011).  The last two rows show what the albedo ($p_{v}$) and individual radii ($r$) of the two components of 2007 TY430 would be if the objects had a reasonable density of $\rho=0.75$ g/cm$^{3}$.}
%\tablenotetext{a}{These objects could be labeled as dwarf planets since their radii are likely larger than 200 km assuming a moderate or lower albedo.}
\end{deluxetable}
\end{center}

%\end{document}             % End of document.

\newpage

%\documentstyle [aj_pt4]{article}    % Specifies the document style.

%\begin{document}

\begin{center}
\begin{deluxetable}{lcccccccc}
%\small
\tablenum{5}
\tablewidth{7 in}
\tablecaption{Equal-sized TNO Binaries}
\tablecolumns{9}
\tablehead{
\colhead{Name} & \colhead{H} & \colhead{$i$}  & \colhead{$e$} & \colhead{$a$} & \colhead{$\Delta$Mag} & \colhead{Sep} & \colhead{$S$} & \colhead{Class} \\ \colhead{} & \colhead{(mag)} & \colhead{(deg)} & \colhead{} & \colhead{(AU)} & \colhead{(mag)} & \colhead{(arcsec)} & \colhead{} & \colhead{} }  
\startdata
                  2003 WU188 &  5.7 &   3.8 & 0.04 & 44.20 &  0.7 &    0.042 &    ...          &  cl          \nl
(60621)           2000 FE8   &  6.9 &   5.9 & 0.40 & 55.02 &  0.6 &    0.044 &    10.2         &  r 5:2       \nl
                  2001 RZ143 &  6.4 &   2.1 & 0.07 & 44.34 &  0.1 &    0.046 &    ...          &  cl            \nl
                  1998 WV24  &  7.5 &   1.5 & 0.04 & 39.14 &  0.3 &    0.051 &    9.4          &  inner cl     \nl
                  2003 QA91  &  5.5 &   2.4 & 0.07 & 44.62 &  0.1 &    0.056 &    ...          &  cl            \nl
(80806)           2000 CM105 &  6.7 &   3.8 & 0.07 & 42.07 &  0.6 &    0.059 &    32.2         &  cl           \nl
                  2003 QR91  &  6.5 &   3.5 & 0.19 & 46.81 &  0.2 &    0.062 &    ...          &  cl            \nl
                  2001 FL185 &  7.1 &   3.6 & 0.07 & 43.95 &  0.8 &    0.065 &    22.1         &  cl           \nl
                  2000 WT169 &  6.1 &   1.7 & 0.01 & 45.02 &   0.43 &  0.07  &    ...          &  cl          \nl
(79360)           1997 CS29  &  5.3 &   2.2 & 0.01 & 43.80 & 0.09 &    0.07  &    28.5         &  cl           \nl
(60458)           2000 CM114 &  6.6 &  19.7 & 0.40 & 59.20 &  0.57 &   0.074 &    5.4          &  sc      \nl
                  2002 VT130 &  5.8 &   1.2 & 0.03 & 42.60 &  0.44 &   0.08  &    ...          &  cl            \nl
(134860)          2000 OJ67  &  6.1 &   1.1 & 0.02 & 43.06 &  0.8 &    0.08  &    25.9         &  cl           \nl
(123509)          2000 WK183 &  6.6 &   2.0 & 0.05 & 44.52 &  0.4 &    0.080 &    31.9         &  cl           \nl
(65489) Ceto      2003 FX128 &  6.3 &  22.3 & 0.82 & 99.68 &  0.6 &    0.085 &    15.4         &  sc/cent       \nl
                  1999 XY143 &  6.0 &   7.2 & 0.08 & 43.10 &  0.38 &   0.09  &    ...          &  cl            \nl
(275809)          2001 QY297 &  5.6 &   1.5 & 0.08 & 44.02 & 0.42 &    0.091 &    27.2         &  cl            \nl            
                  1999 OJ4   &  7.1 &   4.0 & 0.03 & 38.21 & 0.16 &    0.097 &    28.2         &  inner cl            \nl
                  1999 RT214 &  7.8 &   2.6 & 0.05 & 42.77 & 0.81 &    0.107 &    29.5         &  cl             \nl
                  2001 XR254 &  5.7 &   1.2 & 0.03 & 42.96 & 0.09 &    0.107 &    15.1         &  cl            \nl
                  2006 SF369 &  6.7 &  27.6 & 0.38 & 63.18 &  0.1 &    0.11  &    ...          &  r 3:1        \nl
                  2003 TJ58  &  8.0 &   1.0 & 0.09 & 44.55 & 0.50 &    0.119 &    20.5         &  cl            \nl
                  2001 QQ322 &  6.4 &   4.0 & 0.06 & 44.24 & 0.2  &    0.13  &    ...          &  cl            \nl
                  2001 QC298 &  6.1 &  30.5 & 0.13 & 46.56 & 0.58 &    0.130 &    7.1          &  sc     \nl
(148780) Altjira  2001 UQ18  &  5.7 &   5.2 & 0.06 & 44.43 & 0.7  &    0.177 &    ...          &  cl            \nl
                  2000 CQ114 &  7.0 &   2.7 & 0.12 & 45.94 &  0.4 &    0.178 &    30.9         &  cl           \nl
(58534) Logos     1997 CQ29  &  6.6 &   2.9 & 0.12 & 45.05 &  0.4 &    0.20  &    ...          &  cl             \nl
(66652) Borasisi  1999 RZ253 &  5.9 &   0.6 & 0.09 & 44.05 &   0.33 &  0.21  &    ...          &  cl          \nl
                  2000 QL251 &  6.6 &   3.7 & 0.22 & 48.14 &   0.05 &  0.25  &    6.6          &  2:1        \nl
(160091)          2000 OL67  &  6.8 &   2.0 & 0.11 & 45.33 &  0.6 &    0.26  &    ...          &  cl            \nl
(119067)          2001 KP76  &  6.6 &   7.2 & 0.19 & 43.49 &  0.1 &    0.29  &    ...          &  cl            \nl
                  2003 QY90  &  6.4 &   3.8 & 0.05 & 42.96 &  0.1 &    0.34  &    28.2         &  cl              \nl
                  2004 KH19  &  ... &  35.3 & 0.12 & 40.77 & 0.71 &    0.4   &    ...          &  cl            \nl
                  2006 BR284 &  7.3 &  1.2  & 0.04 & 43.82 &  0.50 &   $>0.5$ &   ...          &  cl            \nl
                  2006 JZ81  &  6.9 &  3.6  & 0.08 & 44.53 &  0.98 &   $>0.5$ &   ...          &  cl            \nl
                  2002 XH91  &  5.5 &   5.0 & 0.09 & 44.00 & 1.04* &   0.58  &    ...          &  cl            \nl
(88611)           2001 QT297 &  5.8 &   2.6 & 0.02 & 44.25 &  0.7 &    0.61  &    ...          &  cl            \nl
                  2006 CH69  &  ... &   1.8 & 0.04 & 45.74 & 0.28 &    0.61  &    ...          &  cl            \nl
                  \textbf{2007 TY430} &  \textbf{6.9} &  \textbf{11.3} & \textbf{0.27} & \textbf{39.55} & \textbf{0.1} &    \textbf{0.67}  &    \textbf{36.0}         &  \textbf{r 3:2}        \nl
                  2002 VF130 &  7.2 &  19.5 & 0.12 & 46.04 & 0.31 &    0.733 &    ...          &  sc           \nl
(160256)          2002 PD149 &  6.3 &   4.9 & 0.06 & 43.19 & 0.4  &    0.74  &    ...          &  cl            \nl
                  2000 CF105 &  6.9 &   0.5 & 0.04 & 43.78 &  0.7 &    0.78  &    22.4         &  cl            \nl
                  1998 WW31  &  6.1 &   6.8 & 0.09 & 44.83 &  0.4 &    1.2   &    6.8          &  cl            \nl
                  2003 UN284 &  7.4 &   3.1 & 0.01 & 42.80 &  0.6 &    2.0   &    ...          &  cl            \nl
                  2005 EO304 &  6.3 &   3.4 & 0.07 & 45.45 & 1.2* &    2.67  &    30.9         &  cl            \nl
                  2001 QW322 &  7.8 &   4.8 & 0.02 & 44.18 & 0.0  &    4     &    8.6          &  cl            \nl
\enddata 
\tablecomments{Equal-sized binaries have $\Delta$Mag less than 1 mag.
  The heliocentric orbital elements are from the Minor Planet Center
  and are the semimajor axis ($a$), inclination ($i$), and
  eccentricity ($e$).  H is the absolute magnitude of the object in
  the V-band, $\Delta$Mag is the difference in magnitude of the two
  components, Sep is the discovery seperation of the two components,
  $S$ is the spectral gradient for the object and Class is the
  dynamical class of the object where cl=classical, sc=scattered,
  cent=centaur and r=resonance.  Both 2002 XH91 and 2005 EO304 have
  greater than $\Delta$Mag of 1 mag, but have been included because
  of their large seperations and thus are denoted by a *.  Color information references: Tegler
  and Romanishin (2000), Hainaut and Delsanti (2002), Doressoundiram
  et al. 2002, Tegler and Romanishin (2003), Petit et al. (2008),
  Benecchi et al. (2009). Binary information references: Veillet et
  al. (2002), Noll et al. (2002, 2004a, 2004b, 2008a), Osip et
  al. (2003), Stephens and Noll (2006), Kern and Elliot (2006b),
  Grundy et al. (2007), Petit et al. (2008), Lin et al. (2010), Grundy
  et al. (2011) and Parker et al. (2011).}
%\tablenotetext{a}{These objects could be labeled as dwarf planets since their radii are likely larger than 200 km assuming a moderate or lower albedo.}
\end{deluxetable}
\end{center}

%\end{document}             % End of document.

\newpage

\newpage

\newpage

\newpage

\newpage

\newpage

%\newpage
%
%\begin{figure}
%\epsscale{0.4}
%\centerline{\includegraphics[angle=0,totalheight=0.6\textheight]{rhoalbplot.ps}}
%\caption{The possible density versus albedo for 2007 TY430.  In order
%  to have a reasonable density ($\rho > 0.5$ g/cm$^{3}$) the albedo is
%  required to be at least moderate ($p>0.17$).}
%\label{fig:densityalbedo} 
%\end{figure}

\newpage

\newpage

\newpage

\newpage

%\newpage

%\begin{figure}
%\epsscale{0.4}
%\centerline{\includegraphics[angle=90,totalheight=0.6\textheight]{binarymutualei.ps}}
%\caption{The eccentricity versus the inclinations of the known
%  KBO binaries.  There appears to be no obvious correlations between these
%  parameters.}
%\label{fig:binarymutualei} 
%\end{figure}

\end{document}